\documentclass[11pt,a4paper]{article}
\pdfoutput=1

\usepackage{jheppub}
\usepackage{latexsym}
\usepackage{revsymb}
\usepackage{multirow}
\usepackage{color}
\usepackage[usenames,dvipsnames,svgnames,table]{xcolor}

\usepackage{graphicx}
\usepackage{epsfig}  
\usepackage{epsf}    
\usepackage{dcolumn}
\usepackage{bm}
\usepackage{dcolumn}
\usepackage{textcomp}
\usepackage{float}
\usepackage{hypcap}
\usepackage[]{hyperref}
\usepackage{adjustbox}
\usepackage{multirow}

\usepackage{subfigure}
\usepackage{alphalph}
\usepackage{morefloats}
\usepackage{textcomp}

\newcommand{\be}{\begin{equation}}
\newcommand{\ee}{\end{equation}}
\newcommand{\ba}{\begin{eqnarray}}
\newcommand{\ea}{\end{eqnarray}}
\def\nue{{\nu_e}}
\def\anue{{\bar\nu_e}}

\def\nue{{\nu_e}}
\def\anue{{\bar\nu_e}}

\preprint{IP/BBSR/2016-3}

\title{Physics Reach of DUNE with a Light Sterile Neutrino} 

\author[a]{Sanjib Kumar Agarwalla,}
\author[a]{Sabya Sachi Chatterjee,}
\author[b,c]{Antonio Palazzo}

\affiliation[a]{Institute of Physics, Sachivalaya Marg, Sainik School Post, Bhubaneswar 751005, India}
\affiliation[b]{Dipartimento Interateneo di Fisica ``Michelangelo Merlin'',
Via Amendola 173, 70126 Bari, Italy} 

\affiliation[c]{Istituto Nazionale di Fisica Nucleare (INFN), Sezione di Bari, Via E.\ Orabona 4, I-70126 Bari, Italy}

\emailAdd{sanjib@iopb.res.in}
\emailAdd{sabya@iopb.res.in}
\emailAdd{palazzo@ba.infn.it}

\abstract{
We investigate the implications of one light eV scale sterile neutrino on the
physics potential of the proposed long-baseline experiment DUNE. 
If the future short-baseline experiments confirm the existence of sterile
neutrinos, then it can affect the mass hierarchy (MH) and CP-violation (CPV) 
searches at DUNE. The MH sensitivity still remains above 5$\sigma$ 
if the three new mixing angles ($\theta_{14}, \theta_{24}, \theta_{34}$) 
are all close to $\theta_{13}$. In contrast, it can decrease to 4$\sigma$ 
if the least constrained mixing angle $\theta_{34}$ is close to its 
upper limit $\sim 30^0$. We also assess the sensitivity to the CPV 
induced both by the standard CP-phase $\delta_{13} \equiv \delta$, and 
the new CP-phases $\delta_{14}$ and $\delta_{34}$. In the 3+1 scheme, the
discovery potential of CPV induced by $\delta_{13}$ gets deteriorated
compared to the 3$\nu$ case. In particular, the maximal sensitivity 
(reached around $\delta_{13}$ $\sim$ $\pm$ $90^0$) decreases from 
$5\sigma$ to $4\sigma$ if all the three new mixing angles are close 
to $\theta_{13}$. It can further diminish to almost  $3\sigma$ if 
$\theta_{34}$ is large ($\sim 30^0$). The sensitivity to the CPV 
due to $\delta_{14}$ can reach 3$\sigma$ for an appreciable 
fraction of its true values. Interestingly, $\theta_{34}$ and its 
associated phase $\delta_{34}$ can influence both the 
$\nu_e$ appearance and $\nu_\mu$ disappearance channels 
via matter effects, which in DUNE are pronounced. Hence, DUNE 
can also probe CPV induced by $\delta_{34}$ provided $\theta_{34}$ 
is large. We also reconstruct the two phases $\delta_{13}$ and $\delta_{14}$. 
The typical 1$\sigma$ uncertainty on $\delta_{13}$ ($\delta_{14}$) 
is $\sim20^0$ ($30^0$) if $\theta_{34} =0$. The reconstruction of 
$\delta_{14}$ (but not that of $\delta_{13}$) degrades if $\theta_{34}$ is large.}

\arxivnumber{1603.03759}
\begin{document}
\maketitle

\section{Introduction and Motivation}
\label{introduction}

The high energy physics community is devoting enormous effort at the cosmic, 
intensity, and energy frontiers to unravel the deep long-standing mysteries 
of the Universe. At the forefront of the intensity frontier, ambitious next-generation, 
long-baseline (LBL) neutrino oscillation experiments are under 
consideration~\cite{Pascoli:2013wca,Agarwalla:2013hma,Agarwalla:2014fva,Feldman:2012qt,Stanco:2015ejj},    
whose mission will be that of deciphering the fundamental properties of these elusive particles, 
and of the interactions to which they participate. The main objectives of the LBL experiments 
are to determine the leptonic CP-violation (CPV), and to identify the neutrino mass hierarchy (MH). 
However, the actual scope of these projects is much broader and to some extent unknown. 
As a matter of fact, these setups are potentially able to unveil physics beyond the Standard Model, 
which may be disguised in the form of new neutrino interactions or states.

The LBL experiments are high-precision neutrino interferometers, 
and as such, they are the best place to study CPV, whose manifestation 
is intimately related to interference phenomena. This is clearly the case 
of the standard 3-flavor CPV, which is observable through the interference 
of the oscillations driven by the two distinct (atmospheric and solar) frequencies. 
Several non-standard neutrino properties are sources of additional CPV, 
and the LBL machines would naturally have a pivotal role also in these searches. 
Interestingly, as first evidenced  in~\cite{Klop:2014ima}, the LBL experiments 
may be sensitive to the new CP-phases arising in the presence 
of a light sterile neutrino state. 

Sterile neutrinos represent one of the simplest and most ``innocuous'' 
extensions of the Standard Model and naturally appear in many mechanisms 
of generation of the neutrino masses. Intriguingly, a series of anomalies 
recorded at the short baseline experiments, lends support to the existence of 
such new neutrino states (see~\cite{Abazajian:2012ys,Palazzo:2013me,Gariazzo:2015rra} 
for a review of the topic). These phenomena cannot be accommodated in the 3-flavor 
framework, and require the existence of a fourth mostly sterile mass eigenstate 
at the scale of $\sim$ 1\,eV.

A rich and diverse program of new short-baseline (SBL) experiments is underway to test
such a hypothesis (see the review in~\cite{Lasserre:2014ita}). The SBL experiments
are surely the best place where to look for light sterile neutrinos. In fact, 
at the short baselines, the characteristic oscillating behavior driven by the 
new mass-squared splitting $\Delta m^2_{new} \sim 1$\,eV$^2$ should show up. 
On the other hand, the SBL experiments cannot probe all the properties of
sterile neutrinos. In particular, they are not sensitive to 
the new CPV%
\footnote{In the 3+1 scheme with one light eV-scale sterile neutrino, the oscillation picture at the SBL 
experiments essentially boils down to the 2-flavor scheme, and therefore, one cannot observe CPV. 
When more sterile neutrinos come into play the SBL experiments can be sensitive to CPV.
However, the SBL experiments can access only a restricted number of the CP-phases involved in the model,
while the LBL ones are sensitive to all of them. For example, in the 3+2 scheme, the SBL experiments
are sensitive to one CP-phase over a total of five CP-phases.}
phenomena implied by the sterile neutrinos, which 
need long distances to develop at a detectable level. 
Therefore, in the eventuality of a discovery of a light sterile neutrino, 
the LBL setups would play a complementary role to the SBL experiments, 
due to their unique sensitivity to the new CP-phases.

An unavoidable (less desired) byproduct of this circumstance is that the 
performance of the LBL machines towards their original targets tends 
to be deteriorated in the presence of sterile neutrinos. This is evident for the CPV, 
since the new CP-phases tend to be degenerate with the standard CP-phase $\delta$. 
In addition, the well-known degeneracy between MH and CPV, already existing 
in the 3-flavor scheme, is exacerbated in the presence of the new CP-phases. 
Consequently, the MH discovery potential can also be reduced. Therefore, in view 
of the potential discovery of a light eV-scale sterile neutrino at the SBL experiments, 
a reassessment of the performance of the planned LBL projects appears to be mandatory.

The 4-flavor analyses
\footnote{Such analyses are performed for realistic  values of the new 
active-sterile mixing angles as indicated by the global 3+1 fits~\cite{Giunti:2013aea, Kopp:2013vaa}.}
\cite{Klop:2014ima,Palazzo:2015gja} of the latest data collected 
by the two currently running LBL experiments T2K and NO$\nu$A
\footnote{In principle (see the 4-flavor analysis performed in~\cite{Palazzo:2015wea}), 
the new CP-phases can impact also the $\nu_\mu \to \nu_e$ searches of 
ICARUS~\cite{Antonello:2012pq,Antonello:2015jxa} and OPERA~\cite{Agafonova:2013xsk}. 
However, the very low statistics prevents to extract any information on the CP-phases.} 
have given already some weak hints on the new CP-phases involved in the 3+1 scheme,
and have also evidenced the fragility of the present (weak) indication in favor of the 
normal hierarchy, which emerges in the standard 3-flavor framework. 
In addition, in the recent prospective study~\cite{Agarwalla:2016mrc} of T2K and NO$\nu$A, 
it has been pointed out that even considering the full exposure regime for such two experiments, 
their sensitivity to the standard CP-phase $\delta$, and their discovery potential 
of the neutrino mass hierarchy both get deteriorated in the presence of a sterile neutrino. 
Therefore, the next natural step is to investigate the impact of a sterile neutrino at the more 
sensitive next-generation LBL facilities. In the present work, we focus on the proposed 
Deep Underground Neutrino Experiment (DUNE), which is aiming to achieve new heights 
in the intensity frontier by shooting a very powerful neutrino beam from Fermilab towards 
the Homestake Mine in South Dakota over a distance of 1300 km%
\footnote{The CERN-Pyh\"asalmi baseline of 2290 km actively studied under the umbrella 
of the LBNO collaboration~\cite{Agarwalla:2011hh,Agarwalla:2013kaa} can also be very sensitive 
to these issues. The same is also true for the future T2HK experiment~\cite{Abe:2014oxa,Abe:2015zbg}, 
which is a bigger version of T2K.}.
Our work complements other recent studies performed about DUNE~\cite{Hollander:2014iha,Berryman:2015nua,Gandhi:2015xza}. 
Previous studies on sterile neutrinos at LBL experiments can be found 
in~\cite{Donini:2001xy,Donini:2001xp,Donini:2007yf,Dighe:2007uf,Donini:2008wz,Yasuda:2010rj,Meloni:2010zr,Bhattacharya:2011ee,Donini:2012tt}. 

The paper is organized as follows. In section~\ref{sec:probability}, we introduce the theoretical
framework and present a detailed discussion of the behavior of the 4-flavor $\nu_\mu \to \nu_e$ 
transition probabilities in vacuum and in matter.  In section~\ref{sec:DUNE}, we describe the details 
of the DUNE setup and of the statistical analysis. In section~\ref{sec:MH}, we make some important 
considerations at the level of the bi-events plots and at the level of the energy spectrum, and then 
we present the results of the sensitivity study of the mass hierarchy in the 3+1 scheme. 
Section~\ref{sec:CPV} is devoted to the results concerning the CPV discovery potential 
and the capability of reconstruction of the CP-phases. Finally, we draw our conclusions
in section~\ref{Conclusions}. 

\section{Transition probability in the 3+1 scheme}
\label{sec:probability}

\subsection{Theoretical framework}

In the 3+1 scheme, there is one sterile neutrino $\nu_s$ and the flavor basis
is connected to the mass one through a $4\times4$ matrix, which
can be conveniently parametrized as follows
\begin{equation}
\label{eq:U}
U =   \tilde R_{34}  R_{24} \tilde R_{14} R_{23} \tilde R_{13} R_{12}\,, 
\end{equation} 
where $R_{ij}$ ($\tilde R_{ij}$) is a real (complex) $4\times4$ rotation in the ($i,j$) plane
containing the $2\times2$ matrix 
\begin{eqnarray}
\label{eq:R_ij_2dim}
     R^{2\times2}_{ij} =
    \begin{pmatrix}
         c_{ij} &  s_{ij}  \\
         - s_{ij}  &  c_{ij}
    \end{pmatrix}
\,\,\,\,\,\,\,   
     \tilde R^{2\times2}_{ij} =
    \begin{pmatrix}
         c_{ij} &  \tilde s_{ij}  \\
         - \tilde s_{ij}^*  &  c_{ij}
    \end{pmatrix}
\,,    
\end{eqnarray}
in the  $(i,j)$ sub-block, where we have defined
\begin{eqnarray}
 c_{ij} \equiv \cos \theta_{ij} \qquad s_{ij} \equiv \sin \theta_{ij}\qquad  \tilde s_{ij} \equiv s_{ij} e^{-i\delta_{ij}}.
\end{eqnarray}
The  parameterization in Eq.~(\ref{eq:U}) enjoys the following properties: i) When the mixing
invoving the fourth state is zero $(\theta_{14} = \theta_{24} = \theta_{34} =0)$ 
it returns the 3$\nu$ matrix in its common parameterization.
ii)  For small values of $\theta_{13}$ and of the mixing angles 
involving $\nu_4$, one has $|U_{e3}|^2 \simeq s^2_{13}$, $|U_{e4}|^2 = s^2_{14}$, 
$|U_{\mu4}|^2  \simeq s^2_{24}$ and $|U_{\tau4}|^2 \simeq s^2_{34}$, 
with a clear physical interpretation of the new mixing angles. 
iii) The leftmost positioning of the matrix $\tilde R_{34}$ guarantees that 
the vacuum $\nu_{\mu} \to \nu_{e}$ transition probability is independent of $\theta_{34}$ and of the related CP-phase $\delta_{34}$ (see~\cite{Klop:2014ima}). 
\subsection{Transition probability in Vacuum}
\label{subsec:vacuum}

In the DUNE setup the matter effects play an important role. However, for the
sake of clearness, before discussing the flavor conversion in the
presence of matter we give a brief sketch of the basic formulae in vacuum.
This will help the reader to grasp some new qualitative features, which emerge solely
in the presence of matter. As shown in~\cite{Klop:2014ima}, the $\nu_{\mu} \to \nu_{e}$ 
conversion probability can be written as the sum of three contributions
\begin{eqnarray}
\label{eq:Pme_4nu_3_terms}
P^{4\nu}_{\mu e}  \simeq  P^{\rm{ATM}} + P^{\rm {INT}}_{\rm I}+   P^{\rm {INT}}_{\rm II}\,.
\end{eqnarray}
The first term is positive-definite and is related to the atmospheric mass-squared
splitting. It gives  the leading contribution to the probability and can be written as
\begin{eqnarray}
\label{eq:Pme_atm}
 &\!\! \!\! \!\! \!\! \!\! \!\! \!\!  P^{\rm {ATM}} &\!\! \simeq\,  4 s_{23}^2 s^2_{13}  \sin^2{\Delta}\,,
 \end{eqnarray}
where $\Delta \equiv  \Delta m^2_{31}L/4E$ is the atmospheric oscillating factor, which depends
on the baseline $L$ and the neutrino energy $E$. The second and third contributions
in Eq.~(\ref{eq:Pme_4nu_3_terms}) are interference terms and, as such, they can assume
both positive and negative values. The second term is driven by the standard 
solar-atmospheric interference and can be approximately expressed as
\begin{eqnarray}
 \label{eq:Pme_int_1}
 &\!\! \!\! \!\! \!\! \!\! \!\! \!\! \!\! P^{\rm {INT}}_{\rm I} &\!\!  \simeq\,   8 s_{13} s_{12} c_{12} s_{23} c_{23} (\alpha \Delta)\sin \Delta \cos({\Delta + \delta_{13}})\,.
\end{eqnarray}
The third term is related to the atmospheric-sterile interference and has the following 
form~\cite{Klop:2014ima} 
\begin{eqnarray}
 \label{eq:Pme_int_2}
 &\!\! \!\! \!\! \!\! \!\! \!\! \!\! \!\! P^{\rm {INT}}_{\rm II} &\!\!  \simeq\,   4 s_{14} s_{24} s_{13} s_{23} \sin\Delta \sin (\Delta + \delta_{13} - \delta_{14})\,.
\end{eqnarray}
The transition probability depends on the three small mixing angles $\theta_{13}$, $\theta_{14}$, 
$\theta_{24}$, whose best estimates, derived from the global 3-flavor analyses~\cite{Capozzi:2016rtj,Forero:2014bxa,Gonzalez-Garcia:2014bfa} (concerning $\theta_{13}$) and from the 3+1 
fits~\cite{Giunti:2013aea, Kopp:2013vaa} (concerning $\theta_{14}$ and $\theta_{24}$), turn out to be very similar and we have approximately $s_{13} \sim s_{14} \sim s_{24} \sim 0.15$ (see table~\ref{tab:benchmark-parameters}). Therefore, it is meaningful to treat all such three mixing angles as small quantities of the same order $\epsilon$. Another small quantity entering the transition probability is the ratio of the solar over the atmospheric mass-squared splitting  $\alpha \equiv \Delta m^2_{21}/ \Delta m^2_{31} \simeq \pm 0.03$,  which can be assumed to be of order $\epsilon^2$. From 
Eqs.~(\ref{eq:Pme_atm})-(\ref{eq:Pme_int_2}) we see that the first leading leading term is
of the second order, while the two interference terms  are both of the third order. Hence,
the amplitudes of the two interference terms can be similar.

Before closing this subsection, we would like to highlight the fact that the transition probability in vacuum does not depend on the third mixing angle $\theta_{34}$ and on the associated CP-phase $\delta_{34}$. As we will discuss in the next subsection, this is no more true in the presence of matter, where such a dependency  emerges.

\subsection{Transition probability in Matter}
\label{subsec:Matter}

In the presence of matter, the Hamiltonian in the flavor basis can be written as
\begin{equation}
    H = U K U^\dagger + V \,,
    \label{eq:Hf}
\end{equation}
where $K$ denotes the diagonal matrix containing the wave numbers  
\begin{equation}
K = \mathrm{diag}(0,\, k_ {21},\, k_{31},\, k_{41})\,,
\end{equation}
with $k_{i1} = \Delta m^2_{i1} /2E \, (i=2,3,4)$ and $V$ is the matrix encoding the matter potential
\begin{equation}
\label{eq:V_matrix}
  V=\mathrm{diag}(V_{CC},\,0,\,0,\, -V_{NC})\,,
 \end{equation}
where
\begin{eqnarray}
V_{CC}
&=&
\sqrt{2} \, G_F \, N_e\,
\label{VCC}
\end{eqnarray}
is the charged-current interaction potential of the electron neutrinos with 
the background electrons having number density $N_e$, and
\begin{eqnarray}
V_{NC}
&=&
- \frac{1}{2} \sqrt{2} G_F N_n\,
\label{VNC}
\end{eqnarray}
is the neutral-current interaction potential (common
to all the active neutrino species) with the background 
neutrons having number density $N_n$. For later convenience, we  
also introduce the positive-definite ratio
\begin{equation}
r = - \frac{V_{NC}} {V_{CC}}  = \frac{1}{2} \frac {N_n}{N_e}\,,
\end{equation}
which in the Earth  crust is $r \simeq 0.5$.
In order to simplify the treatment of matter effects, it is useful to introduce the new basis
\begin{equation}
\bar \nu = \bar U{^\dag} \nu\,,
\label{eq:newbasis}
\end{equation}
where
\begin{equation}
\bar U = \tilde R_{34} R_{24} \tilde R_{14}\,
\end{equation}
is the part of the mixing matrix defined in Eq.~(\ref{eq:U}) that contains only the rotations involving the fourth neutrino mass eigenstate. The mixing matrix $U$ is split as follows  
\begin{equation}
U = \bar U U_{3\nu}\,
\end{equation}
where $U_{3\nu}$ is the $4\times4$ matrix which contains the standard 3-flavor mixing
matrix in the (1,2,3) sub-block. In the new basis, the Hamiltonian takes the form
\begin{equation} \label{eq:Hbar}
   \bar H = \bar H^{\rm {kin}} + \bar H^{\rm {dyn}} = U_{3\nu} K U_{3\nu}^\dagger
    + {\bar U}^\dagger V {\bar U} \,,
\end{equation}
where the first term is the  kinematic contribution describing the
oscillations in vacuum, and the second one represents a nonstandard dynamical term.
As shown in~\cite{Klop:2014ima} (see the appendix), since $|k_{41}|$ is much bigger than one 
and much bigger than $|k_{21}|$ and $|k_{31}|$, one can reduce the dynamics to 
an effective 3-flavor system. Indeed, from Eq.~(\ref{eq:Hbar}) one has that the (4,4) entry 
of $\bar H$ is much bigger than all the other elements and
 the fourth eigenvalue of  $\bar H$ is much larger than
the other three ones. As a result, the state $\bar \nu_s$ evolves independently of the others.
Extracting the submatrix with indices  $(1,2,3)$ from  $\bar H$, one obtains the $3\times 3$ 
Hamiltonian
\begin{equation} \label{eq:Hbar_3nu}
   \bar H_{3\nu} = \bar H_{3\nu}^{\rm {kin}} + \bar H_{3\nu}^{\rm {dyn}}\,
 \end{equation}
governing the evolution of the $(\bar \nu_{e}, \bar \nu_{\mu}, \bar \nu_{\tau})$ system,
whose dynamical part has the form~\cite{Klop:2014ima} 
\begin{eqnarray} \label{eq:Hdyn_1}
\footnotesize
\arraycolsep=3pt
\medmuskip = 1mu
    \bar H^{\rm {dyn}}_{3\nu} = 
   V_{CC}   \,  \begin{bmatrix}
	|\bar U_{e1}|^2 + r|\bar{U}_{s1}|^2 & r\bar{U}_{s1}^* \bar{U}_{s2} & r\bar{U}_{s1}^* \bar{U}_{s3}
	\\
	\dagger & r|\tilde{U}_{s2}|^2
	& r\bar{U}_{s2}^* \bar{U}_{s3}
	\\
	\dagger & \dagger
	& r|\bar{U}_{s3}|^2
    \end{bmatrix} \,,
\end{eqnarray}
where we have indicated with $\dagger$ the complex conjugate of the
element with the same two indices inverted. In deriving Eq.~(\ref{eq:Hdyn_1}) we have used the relations $\bar U_{e2} = \bar U_{e3} = \bar U_{\mu 3} = 0$. Considering the explicit expressions
of the elements of $\bar U$, and taking their first-order expansion in the small
mixing angles $\theta_{i4}$ ($i = 1,2,3$), Eq.~(\ref{eq:Hdyn_1}) takes the form
\begin{eqnarray} \label{eq:Hdyn_2}
\footnotesize
\arraycolsep=3pt
\medmuskip = 1mu
    \bar H^{\rm{dyn}}_{3\nu}  \approx  
   V_{\rm{CC}} \!   \begin{bmatrix}
	1 - (1-r) s^2_{14}  & r  \tilde s_{14} s_{24}  & r  \tilde s_{14} \tilde s_{34}^*
	\\
	\dagger & r  s_{24}^2 & r s_{24} \tilde s_{34}^* 
	\\
	\dagger & \dagger
	& r s_{34}^2
    \end{bmatrix} \,.
\end{eqnarray}
From Eq.~(\ref{eq:Hdyn_2}) we see that for vanishing sterile neutrino 
angles ($\theta_{14} = \theta_{24} = \theta_{34} = 0$) one recovers the (diagonal) 
standard 3-flavor MSW Hamiltonian. In general, in the 4-flavor case, the Hamiltonian
in Eq.~(\ref{eq:Hdyn_2}) encodes both diagonal and off-diagonal perturbations.
We can observe that these corrections are formally equivalent to non-standard neutrino 
interactions (NSI)%
\footnote{We stress that this is only a formal analogy. The real NSI are mediated by heavy
particles. In contrast, in the case of sterile neutrinos there is no heavy mediator and the NSI-like structure of the Hamiltonian is merely related to the fact that we are working in the new basis introduced in  Eq.~(\ref{eq:newbasis}),  which is rotated with respect to the original flavor basis and is particularly convenient to handle the problem under consideration. It is worthwhile to mention that a similar analogy has been noticed concerning the solar neutrino transitions in the presence of sterile species~\cite{Palazzo:2011rj}.} 
\begin{eqnarray} \label{eq:Hdyn_3}
\footnotesize
\arraycolsep=3pt
\medmuskip = 1mu
    \bar H^{\rm{dyn}}_{3\nu}  \approx  
   V_{\rm{CC}} \!   \begin{bmatrix}
	1 +\varepsilon_{ee}  &   \varepsilon_{e\mu}  & \varepsilon_{e\tau}
	\\
	\dagger & \varepsilon_{\mu\mu} & \varepsilon_{\mu\tau}
	\\
	\dagger & \dagger
	& \varepsilon_{\tau\tau}
    \end{bmatrix} \,.
\end{eqnarray}
This formal analogy is helpful to qualitatively understand the sensitivity to the new dynamical effects implied by the sterile neutrinos. 
It is well known that in the $\nu_\mu \to \nu_e$ channel, the NSI that play the most important role are $\varepsilon_{e\mu}$ and  
$\varepsilon_{e\tau}$, while in the $\nu_\mu \to \nu_\mu$ channel  $\varepsilon_{\mu\tau}$ has the biggest impact.  
If one assumes that the three new mixing angles have the same value $s_{14}^2 = s_{24}^2 =  s_{34}^2 = 0.025$, 
all the corrections in Eq.~(\ref{eq:Hdyn_3}) are very small ($|\varepsilon_{\alpha\beta}| \simeq 0.01$)  and 
they have a negligible impact. In this case, the dynamics is almost equivalent to that of  the standard 3-flavor case. 
In contrast, if one allows the third mixing angle to assume values close to its upper bound  ($\theta_{34} \sim 30^0$), 
the elements of the third column of the Hamiltonian in Eq.~(\ref{eq:Hdyn_3}) can be appreciably 
larger ($|\varepsilon_{e\tau}| \simeq  |\varepsilon_{\mu\tau}| \simeq 0.04$ and $|\varepsilon_{\tau\tau}| \simeq 0.13$). 
In this last case, one may expect a noticeable impact of $\varepsilon_{e\tau} 
\equiv r \tilde s_{14}\tilde s_{34}^*$ in the $\nu_\mu \to \nu_e$ appearance probability,
and of $\varepsilon_{\mu\tau} \equiv r s_{24}\tilde s_{34}^*$ 
in the $\nu_\mu \to \nu_\mu$ survival probability.  
As a consequence both channels should have some sensitivity to the third CP-phase $\delta_{34}$ since
$\tilde s_{34}^* \equiv s_{34} e^{i\delta_{34}}$. Therefore, in contrast to the vacuum case, in matter, 
the flavor conversion is sensitive to the two parameters $\theta_{34}$ and $\delta_{34}$. 
Concerning the $\nu_\mu \to \nu_e$ appearance probability, this behavior has been already pointed out 
in the analytical treatment presented in~\cite{Klop:2014ima} (see the appendix therein), and successively 
noticed in the numerical simulations performed in~\cite{Gandhi:2015xza}. Regarding the $\nu_\mu \to \nu_\mu$ 
survival probability, to our knowledge the dependency on ($\theta_{34}, \delta_{34}$) has not been noted before 
in the literature. The dependency of both channels from these parameters represents a very interesting feature, 
because in favorable circumstances (i.e., a large value of $\theta_{34}$),  the LBL experiments may be sensitive 
not only to the two CP-phases $\delta_{13}$ and $\delta_{14}$ but also to the CP-phase $\delta_{34}$. 
Hence, the searches performed at DUNE may provide full access to the rich CPV structure of the 3+1 scheme. 
In order to illustrate the peculiar role of the third mixing angle, in our numerical study we will consider the following 
three benchmark cases: $\theta_{34} = 0$,  $\theta_{34} = 9^0$ (i.e., equal to the other two mixing angles 
$\theta_{14}$ and $\theta_{24}$) and $\theta_{34} = 30^0$ (which is a value close to its current upper bound). 
For clarity, we report these values in the second column of Table~\ref{tab:benchmark-parameters}. 

For completeness, we show how to proceed to calculate the oscillation probabilities in matter,
by reducing the dynamics of the 4-flavor system to that of a 3-flavor one~\cite{Klop:2014ima}. 
To this purpose, it is helpful to define the evolution operator, which, in the rotated basis $\bar \nu$, has the form
\begin{equation}
    \bar {S} \equiv e^{-i \bar H L} \approx
    \begin{pmatrix}
   	e^{-i \bar H_{3\nu} L} & \boldsymbol{0} \\
	\boldsymbol{0} & e^{-i k_{41} L}
    \end{pmatrix}\,,
\end{equation}
and is related to the evolution operator in the original flavor basis by the unitary transformation
\begin{equation}
     S = \bar{U} \bar{S} \bar{U}^\dagger \,.
\end{equation}
The $S$ matrix describes the flavor conversion after traversing the distance $L$
\begin{equation}
\nu_\beta(L) = S_{\beta\alpha} \nu_\alpha(0)\,,
\end{equation}
and the 4-flavor oscillation probability is given by%
\footnote{Note the inverted order of the two flavor indexes $\alpha$ and $\beta$ 
in the definitions of the oscillation probability and of the evolution operator.}
\begin{equation}
P^{4\nu}_{\alpha \beta} \equiv P^{4\nu}(\nu_\alpha \to \nu_\beta; L) = |S_{\beta\alpha}|^2\,.
\end{equation}
The block-diagonal form of $\bar{S}$ allows us to express the elements of $S$ in terms
of those of $\bar S$ and thus reduce the 4-flavor problem to a 3-flavor one. In fact, 
exploiting the relations $\bar U_{e2} = \bar U_{e3} = \bar U_{\mu 3} = 0$,
one can easily derive the following expression for the $\nu_\mu \to \nu_e$ transition amplitude
and for that of the  $\nu_\mu \to \nu_\mu$ survival amplitude
\begin{eqnarray}
\label{eq:Sem}
S_{e\mu} &=& \bar U_{e1} \left[\bar {U}_{\mu 1}^*  \bar {S}_{ee} +  \bar {U}_{\mu 2}^* \bar {S}_{e\mu}\right] 
+ \bar U_{e4} \bar U_{\mu4}^*\bar S_{ss}\,, \\
\label{eq:Smm}
S_{\mu\mu} &=& \bar U_{\mu 1}^* \left[\bar {U}_{\mu 1}  \bar {S}_{ee} +  \bar {U}_{\mu 2}\bar {S}_{\mu e}\right]  + \bar U_{\mu 2}^* \left[\bar {U}_{\mu 1}  \bar {S}_{e\mu} +  \bar {U}_{\mu 2}\bar {S}_{\mu \mu}\right] 
	             + |\bar U_{\mu 4}|^2  \bar S_{ss}\,,
\end{eqnarray}
where the five relevant matrix elements of the matrix $\bar U$ are given by
\begin{eqnarray}
\label{eq:pme_gen_mix}
   \bar U_{e 1} &=&  c_{14}\\    
    \nonumber    
    \bar U_{\mu 1} &=&  s_{14} s_{24} e^{i \delta_{14}} \\    
    \nonumber
      \bar U_{\mu 2} &=& c_{24}\\
     \nonumber
       \bar U_{e4}  &=&  s_{14} e^{-i \delta_{14}} \\
     \nonumber
       \bar U_{\mu 4} &=&  c_{14} s_{24} \,.
\end{eqnarray}
We can immediately observe that in vacuum, both $S_{e\mu}$ and $S_{\mu\mu}$ are independent
of the mixing angle $\theta_{34}$ and  the associated CP-phase $\delta_{34}$. In fact, in vacuum, 
the elements of $\bar S$ exactly return the corresponding 3-flavor expressions, which may 
depend solely on the CP-phase $\delta_{13}$. Therefore, in vacuum, both the $\nu_e$ appearance 
and $\nu_\mu$ disappearance oscillation probabilities depend only on the two CP-phases $\delta_{13}$ and $\delta_{14}$. 
In addition, we have verified that the $\nu_\mu$ disappearance probability depends very weakly on such two phases.  
In contrast, in matter, $S_{e\mu}$ and $S_{\mu\mu}$ depend on the mixing angle $\theta_{34}$ and on its associated CP-phase 
$\delta_{34}$  through the elements $\bar S_{ee}$,  $\bar S_{e\mu}$, $\bar S_{\mu e}$ and $\bar S_{\mu\mu}$. 
As discussed above, this is due to the NSI-like structure of the matter term of the Hamiltonian in the 
rotated basis [see Eq.~(\ref{eq:Hdyn_2}) and Eq.~(\ref{eq:Hdyn_3})]. Taking the modulus squared of the 
elements $S_{e\mu}$ and  $S_{\mu\mu}$ in Eqs.~(\ref{eq:Sem})-(\ref{eq:Smm}) one arrives at the expressions 
of the $\nu_e$ appearance and $\nu_\mu$ disappearance probabilities. In the calculation one should take 
into account that $\bar S_{ss} =  e^{-i k_{41} L}$ oscillates very fast,  and that all the associated terms 
are averaged by the energy resolution of the detector.

\begin{table}[t]
\begin{center}
{
\newcommand{\mc}[3]{\multicolumn{#1}{#2}{#3}}
\newcommand{\mr}[3]{\multirow{#1}{#2}{#3}}
\begin{tabular}{|c|c|c|}
\hline\hline
\mr{2}{*}{\bf Parameter} & \mr{2}{*}{\bf True Value} & \mr{2}{*}{\bf Marginalization Range} \\
  & &  \\
\hline\hline
\mr{2}{*}{$\sin^2{\theta_{12}}$} & \mr{2}{*}{0.304} & \mr{2}{*}{Not marginalized} \\
  & &  \\
\hline
\mr{2}{*}{$\sin^22\theta_{13}$} & \mr{2}{*}{$0.085$} & \mr{2}{*}{Not marginalized} \\ 
  & &  \\
\hline
\mr{2}{*}{$\sin^2{\theta_{23}}$} & \mr{2}{*}{0.50} & \mr{2}{*}{[0.34, 0.68]} \\
  & &  \\
\hline
\mr{2}{*}{$\sin^2{\theta_{14}}$} & \mr{2}{*}{0.025} & \mr{2}{*}{Not marginalized} \\
  & &  \\  
\hline
\mr{2}{*}{$\sin^2{\theta_{24}}$} & \mr{2}{*}{0.025} & \mr{2}{*}{Not marginalized} \\
  & &  \\ 
\hline
\mr{2}{*}{$\sin^2{\theta_{34}}$} & \mr{2}{*}{0, 0.025, 0.25} & \mr{2}{*}{Not marginalized} \\
  & &  \\  
\hline
\mr{2}{*}{$\delta_{13}/^{\circ}$} & \mr{2}{*}{[- 180, 180]} & \mr{2}{*}{[- 180, 180]} \\
  & &  \\
\hline
\mr{2}{*}{$\delta_{14}/^{\circ}$} & \mr{2}{*}{[- 180, 180]} & \mr{2}{*}{[- 180, 180]} \\
  & &  \\
\hline
\mr{2}{*}{$\delta_{34}/^{\circ}$} & \mr{2}{*}{[- 180, 180]} & \mr{2}{*}{[- 180, 180]} \\
  & &  \\  
\hline
\mr{2}{*}{$\frac{\Delta{m^2_{21}}}{10^{-5} \, \rm{eV}^2}$} & \mr{2}{*}{7.50} & \mr{2}{*}{Not marginalized} \\
  & &  \\
\hline
\mr{2}{*}{$\frac{\Delta{m^2_{31}}}{10^{-3} \, \rm{eV}^2}$ (NH)} & \mr{2}{*}{2.475} &\mr{2}{*}{Not marginalized} \\
  & & \\
\hline
\mr{2}{*}{$\frac{\Delta{m^2_{31}}}{10^{-3} \, \rm{eV}^2}$ (IH)} & \mr{2}{*}{- 2.4} &\mr{2}{*}{Not marginalized} \\
  & & \\
\hline
\mr{2}{*}{$\frac{\Delta{m^2_{41}}}{\rm{eV}^2}$} & \mr{2}{*}{1.0} & \mr{2}{*}{Not marginalized} \\
  & &  \\
\hline\hline
\end{tabular}
}
\caption{Parameter values/ranges used in the numerical calculations. The second column reports 
the true values of the oscillation parameters used to simulate the ``observed'' data set. The third column reports the range over which $\sin^2\theta_{23}$, and the three CP-phases 
$\delta_{13}$,  $\delta_{14}$  and $\delta_{34}$ are varied while minimizing the $\chi^{2}$ to obtain the final results.}
\label{tab:benchmark-parameters}
\end{center}
\end{table}

\section{Description of DUNE setup and Statistical Analysis}
\label{sec:DUNE}

Hosted at Fermilab, DUNE is a next-generation long-baseline neutrino oscillation experiment
which will play an important role in the future neutrino roadmap to unravel the fundamental properties 
of neutrino~\cite{Acciarri:2016crz,Acciarri:2015uup,Strait:2016mof,Acciarri:2016ooe,Adams:2013qkq}. 
DUNE is a quite ambitious project, and to accomplish its broad and rich physics objectives as a world-class facility, 
it is needed to develop three major components: a) an intense ($\sim$ megawatt), 
wide-band neutrino beam at Fermilab, b) a fine-grained, high-precision near neutrino detector just 
downstream of the neutrino source, and c) a massive ($\sim$ 40 kt) liquid argon time-projection chamber (LArTPC) 
far detector housed deep underground at the Sanford Underground Research Facility (SURF) 1300 km away in Lead, South Dakota. 
The LArTPC technology is quite effective for uniform and high accuracy imaging of massive active volumes~\cite{Amerio:2004ze}.
One of the merits of a LArTPC is that it works like a totally active calorimeter where one can 
detect the energy deposited by all final-state particles, providing an excellent energy resolution
over a wide range of energies inevitable to study the first and second oscillation maxima.
We assume a fiducial mass of 35 kt for the far detector in our simulation, and consider the
detector properties which are given in Table 1 of Ref. \cite{Agarwalla:2011hh}. As far as the 
neutrino beam specifications are concerned, we consider a proton beam power of 708 kW
in its initial phase with a proton energy of 120 GeV which can deliver $6 \times 10^{20}$
protons on target in 230 days per calendar year. In our simulation, we have used the fluxes
which were estimated assuming a decay pipe length of 200 m and 200 kA horn current~\cite{mbishai}.
We consider a total run time of ten years for this experiment which is equivalent 
to a total exposure of 248 kt $\cdot$ MW $\cdot$ year. We assume that the DUNE experiment 
would use half of its full exposure in the neutrino mode, and the remaining half would be used 
during antineutrino run. In our simulation, we take the reconstructed neutrino and 
anti-neutrino energy range to be 0.5 GeV to 10 GeV. To incorporate the systematic uncertainties, 
we consider an uncorrelated 5\% normalization error on signal, and 5\% normalization error on 
background for both the appearance and disappearance channels to analyze the prospective data 
from the DUNE experiment. We consider the same set of systematics for both the neutrino and 
antineutrino channels which are also uncorrelated. For both $\nue$ and $\anue$ appearance channels, 
the backgrounds mainly arise from three different sources: a) the intrinsic $\nue$/$\anue$ contamination of the beam, 
b) the number of muon events which will be misidentified as electron events, and c) the neutral current events. 
Our assumptions on various components of the DUNE set-up are slightly different compared to what have been 
considered for the Conceptual Design Report (CDR) reference design in Ref.~\cite{Acciarri:2015uup}%
\footnote{But, needless to mention, these assumptions on the beam fluxes, various detector characteristics, 
and systematic uncertainties are preliminary, and are expected to evolve with time as our understanding 
about the key components of the DUNE experiment is going to be refined/improved with the help of ongoing R\&D efforts.}. 
It has been shown in Ref.~\cite{Acciarri:2015uup} that considering the CDR
reference beam design and with an exposure of 1320 kt $\cdot$ MW $\cdot$ year, CPV
can be determined for 75\% of $\delta$ values at 3$\sigma$ confidence level. We have checked 
that similar coverage in $\delta$ values for establishing CPV can be achieved with our 
assumptions on the DUNE setup if we consider the same amount of exposure and the same
oscillation parameters.

The experimental sensitivities presented in this work are estimated using the GLoBES 
software~\cite{Huber:2004ka,Huber:2007ji} along with its new physics tools. 
We incorporate the 4-flavor effects both in the $\nu_\mu \to \nu_e$ appearance channel, 
and in the $\nu_\mu \to \nu_\mu$ disappearance channel. The same is also 
true for the anti-neutrino mode. We take the neutrino interaction cross-sections from 
Refs.~\cite{Messier:1999kj,Paschos:2001np}, where the authors gave the cross-section
for water and isoscalar targets. To obtain cross-sections for LAr, we have scaled the 
inclusive charged current cross-sections of water by a factor of 1.06 for neutrino,
and 0.94 for anti-neutrino~\cite{zeller,petti-zeller}. The details of the statistical method 
that we follow in the present work, are exactly similar to what have been described 
in section 4 of Ref.~\cite{Agarwalla:2016mrc}. The only difference that we have 
in the present paper is that, we have also analyzed the cases where the active-sterile
mixing angle $\theta_{34}$, and its associated CP-phase $\delta_{34}$ are non-zero,
since the impact of these parameters can be noticeable for the DUNE baseline as 
can be seen from our sensitivity results which we present in the next sections.
Table~\ref{tab:benchmark-parameters} shows the true values of the oscillation parameters 
and their marginalization ranges which we consider in our simulation. As far as the mixing 
angles involving the fourth state are concerned ($\theta_{14}$, $\theta_{24}$, and $\theta_{34}$),
we take their values to be fixed as given in Table~\ref{tab:benchmark-parameters} 
while generating the data and also in the fit. For non-zero $\sin^2\theta_{34}$, we take the 
true values of 0.025 and 0.25 which are well within its allowed range~\cite{Giunti:2013aea, Kopp:2013vaa}.
We vary the true values of $\delta_{14}$ and $\delta_{34}$ in their allowed ranges of [$-\pi,\pi$], 
and they have been marginalized over their full ranges in the fit as required. 
We consider the mass-squared splitting $\Delta m^2_{41} = 1\,$eV$^2$, which
is the value currently suggested by the SBL anomalies. However, we stress that our
results would be unchanged for different choices of this parameter, provided 
that  $\Delta m^2_{41} \gtrsim 0.1\,$eV$^2$. For such values, the very fast
oscillations driven by the new large mass-squared splitting are completely averaged
by the finite energy resolution of the detector. For the same reason, 
DUNE is insensitive to sgn($\Delta m^2_{41}$) and we can confine
our study to positive values. For the 2-3 mixing angle, we consider 
the maximal mixing ($\pi/4$) as the true value, and in the fit, we marginalize over the range 
given in table~\ref{tab:benchmark-parameters}. We marginalize over both the choices of hierarchy 
in the fit for all the analyses, except for the mass hierarchy discovery studies where our goal is to 
exclude the wrong hierarchy in the fit. We take the line-averaged constant Earth matter 
density\footnote{The line-averaged constant Earth matter density has been computed using the 
Preliminary Reference Earth Model (PREM)~\cite{PREM:1981}.} of 2.87 g/cm$^{3}$ 
for the DUNE baseline. In our calculation, we do not explicitly consider the near detector
of DUNE which may provide some information on $\theta_{14}$ and $\theta_{24}$, 
but certainly, the near detector data are not sensitive to the CP-phases in which we are 
interested. In our simulation, we have performed a full spectral analysis using
the binned events spectra for the DUNE setup. In the numerical analysis, 
the Poissonian $\Delta\chi^{2}$ is marginalized over the uncorrelated systematic 
uncertainties using the method of pulls as described in Refs.~\cite{Huber:2002mx,Fogli:2002pt}. 
To present the results, we display the $1,2,3\sigma$ confidence levels for 1 d.o.f.
using the relation $\textrm{N}\sigma \equiv \sqrt{\Delta\chi^2}$.
In~\cite{Blennow:2013oma}, it was demonstrated that the above relation is valid in the
frequentist method of hypothesis testing.

\section{Mass hierarchy discovery potential in the 3+1 scheme}
\label{sec:MH}

In this section, we discuss the sensitivity of DUNE to the neutrino mass hierarchy. 
As a first step we provide a description at the level of the signal events aided by
bi-event plots. Second, we highlight the role of the spectral shape information in the
hierarchy discrimination. Finally, we present the results of the full numerical analysis. 

\subsection{Discussion at the level of bi-events plots}
\label{subsec:biprob}

 \begin{figure}[t!]
\centerline{
 \includegraphics[height=7.5 cm,width=7.5cm]{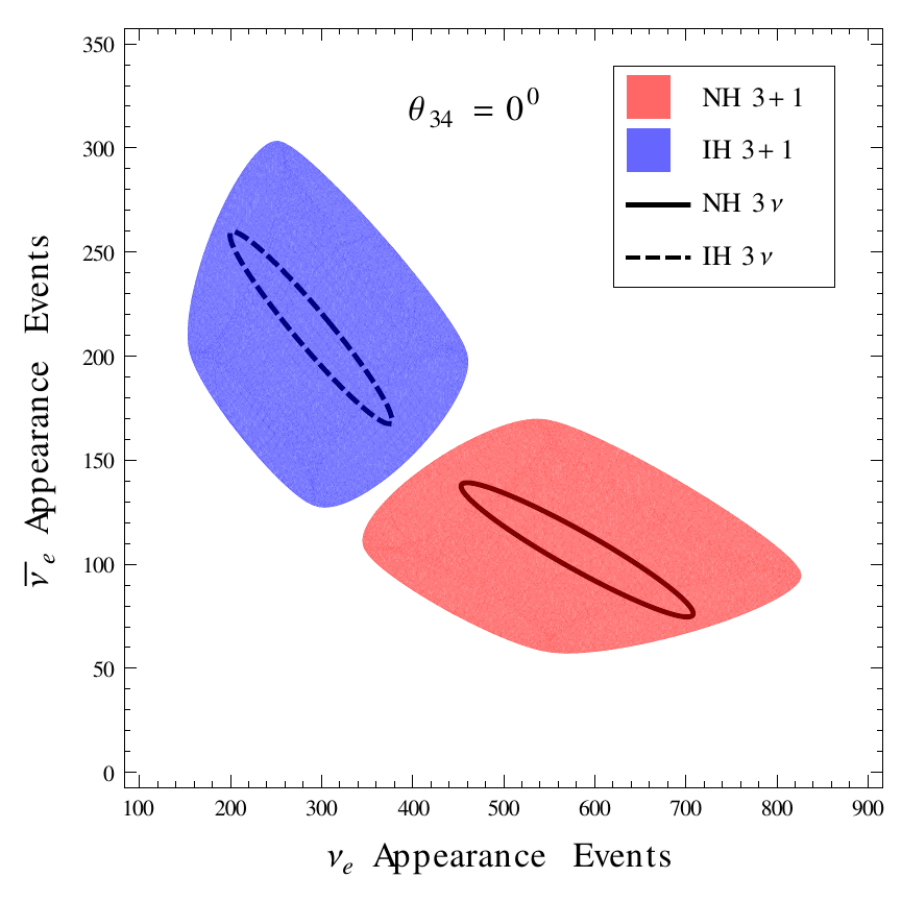}
 \includegraphics[height=7.5 cm,width=7.5cm]{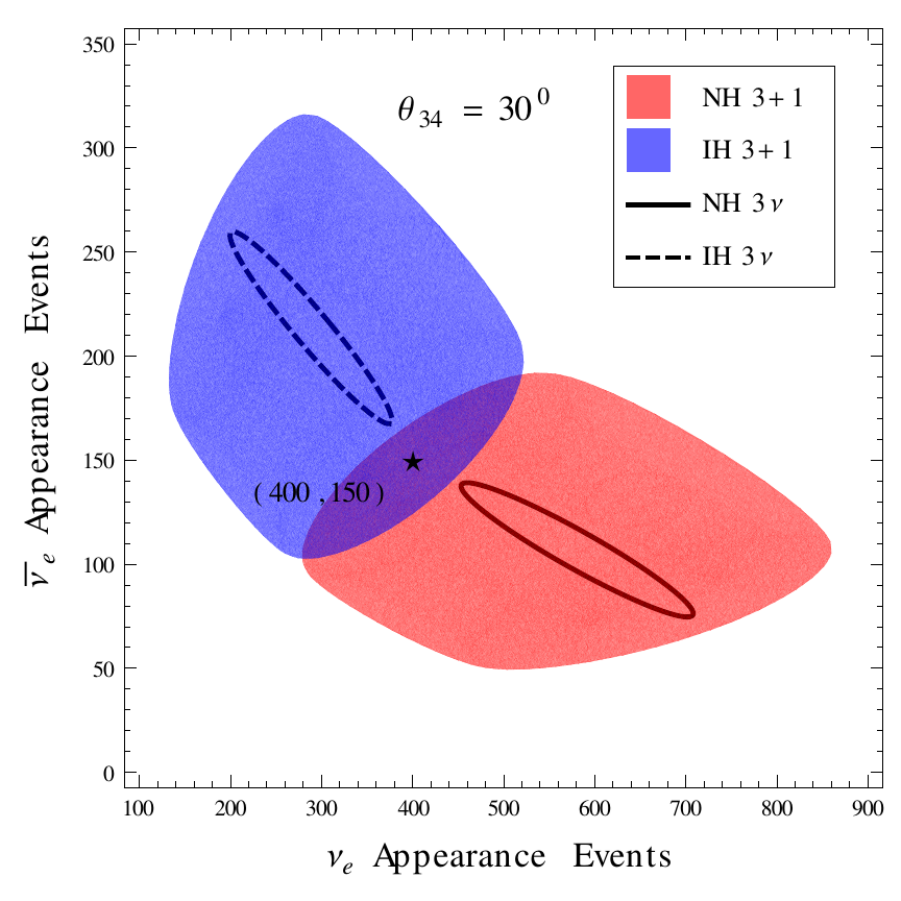}
 }
 \caption{The colored shaded blobs represent the bi-event plots 
for DUNE in the 3+1 scheme. The left panel corresponds to the case $\theta_{34} = 0$,
while the right panel corresponds to the case $\theta_{34} = 30^0$. The blobs in the left panel 
are obtained by varying the CP-phases $\delta_{13}$ and $\delta_{14}$ in the range $[-\pi,\pi]$.
In the right panel also the CP-phase $\delta_{34}$ is allowed to vary in the same range.
The black curves show the 3-flavor ellipses for comparison. In this case, there is
only one running parameter, which is the CP-phase $\delta_{13}$.}
 \label{fig:bievents-conv.pdf}
 \end{figure}

In Fig.~\ref{fig:bievents-conv.pdf}, we introduce the bi-event plots,
in which the two axes represent the total number of $\nu_e$ ($x$-axis)
 and  $\bar \nu_e$ ($y$-axis) events%
\footnote{In the 3+1 scheme, these kind of plots were introduced for the first time 
in Ref.~\cite{Agarwalla:2016mrc} in the context of T2K and NO$\nu$A.}.
In both panels, the 3-flavor case is represented by the black ellipses, which
are obtained varying the CP-phase $\delta_{13}$ in the range $[-\pi,\pi]$.
In the 3+1 scheme, there are more CP-phases and the bi-event plot
becomes a blob. In particular in the left panel, where we have taken $\theta_{34} =0$,
we vary the two phases $\delta_{13}$ and $\delta_{14}$ in the range $[-\pi,\pi]$. In addition, in the right panel,
which corresponds to $\theta_{34} = 30^0$, we also vary the phase $\delta_{34}$
in the same range.
The 3+1 blobs can be seen as a convolution of an infinite ensemble of ellipses
having different orientations or, alternatively, as a scatter plot obtained by varying simultaneously 
the CP-phases $\delta_{13}$, $\delta_{14}$ (and also $\delta_{34}$ in the right panel).
One can see that in both the panels, the separation between the blobs corresponding to the two 
different hierarchies in the 3+1 scheme gets reduced as compared to the 3-flavor ellipses.
In particular, as evident from the right panel, for $\theta_{34} = 30^0$, 
there is a region where the two hierarchies overlap. For each point of this region, 
there are two possible choices of the three CP-phases 
(one choice corresponding to NH and the other one to IH). 
In these cases, there is a complete degeneracy between the two hierarchies 
at the level of the total number of events (both neutrino and antineutrino), 
and the information based on total event rates cannot distinguish between NH and IH. 
The point marked by star in the right panel corresponding to 400 $\nu_e$ and 150 $\bar\nu_e$ 
events for both NH and IH is a representative degenerate case, which we intend to discuss 
in more detail in the following subsection.

\subsection{Role of the energy spectrum}
\label{subsec:MHspec}

\begin{figure}[t!]
\centerline{
\includegraphics[width=\textwidth]{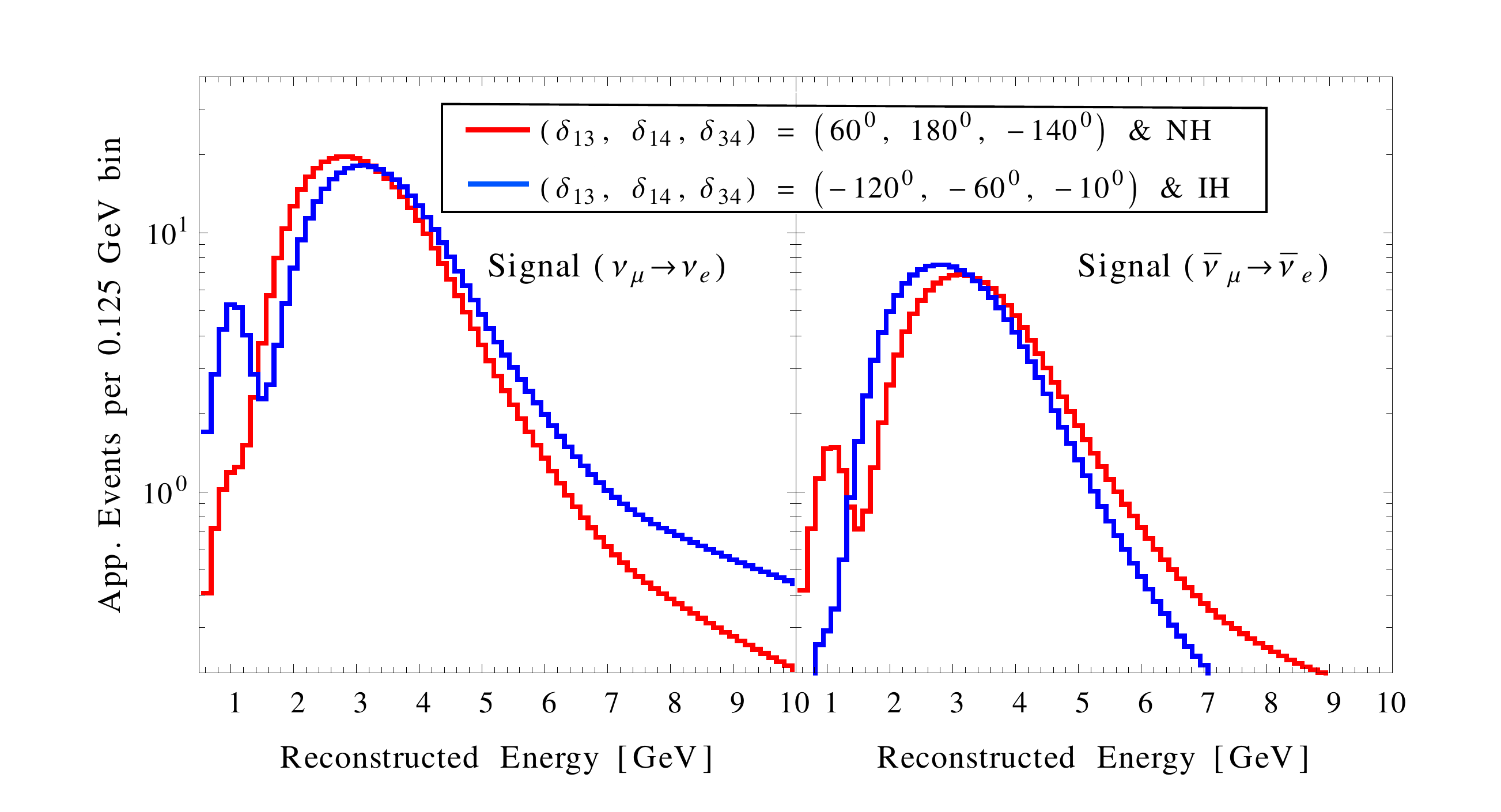}
}
\caption{Expected signal event spectrum of DUNE for the $\nue$ (left panel) and 
$\bar \nu_e$ (right panel) appearance channel as a function of the reconstructed 
neutrino energy. The red (blue) histogram corresponds to the case of NH (IH). 
All the spectra are obtained for $\theta_{34} = 30^0$. For the values of the three CP-phases 
($\delta_{13}, \delta_{14}, \delta_{34}$) indicated in the legend, which are different for NH and IH, 
the number of neutrino and antineutrino events are identical for the two hierarchies. 
This behavior can be visualized in the right panel of the convoluted bi-events plot shown 
in Fig.~\ref{fig:bievents-conv.pdf}, where the two choices of parameters correspond to the 
same point (indicated by a star). In this particular case, the distinction between the two 
mass hierarchies relies solely upon the difference in the shape of the energy spectrum.}
\label{fig:events-spectra}
\end{figure}

The DUNE experiment employs a wide-band energy spectrum, whose shape
brings additional information to that given by the total number of events.
Therefore, even when there is a complete degeneracy at the level of the total number of events,
it is possible to distinguish the two hierarchies at some confidence level. Figure~\ref{fig:events-spectra} 
serves to illustrate this point. It represents the expected signal event spectrum of DUNE for the $\nue$ (left panel) 
and $\bar \nu_e$ (right panel) appearance channels as a function of the reconstructed neutrino energy. 
The red (blue) histogram corresponds to the case of NH (IH). All the spectra are obtained for $\theta_{34} = 30^0$. 
For the values of the three CP-phases ($\delta_{13}, \delta_{14}, \delta_{34}$) indicated in the legend, 
which are different for the two hierarchies, the number of neutrino and antineutrino events are identical for the two hierarchies. 
This fact can be visualized in the right panel of the bi-events plot shown in Fig.~\ref{fig:bievents-conv.pdf}, where 
the two choices of parameters correspond to the same point (indicated by a star). In this particular case, the distinction 
between the two mass hierarchies relies solely upon the difference in the shape of the energy spectrum, 
which, as one can appreciate in Fig.~\ref{fig:events-spectra}, is sizable. Therefore, we expect a good discrimination 
potential also for those points which are completely degenerate at the event level. This will be confirmed by the 
numerical simulation presented in the following, where we find that the pure-shape information is able to guarantee 
a minimal 4$\sigma$ level separation between the two mass hierarchies.

\subsection{Numerical Results}
\label{subsec:MHnum}

\begin{figure}[t!]
\centerline{
\includegraphics[height=8. cm,width=8.cm]{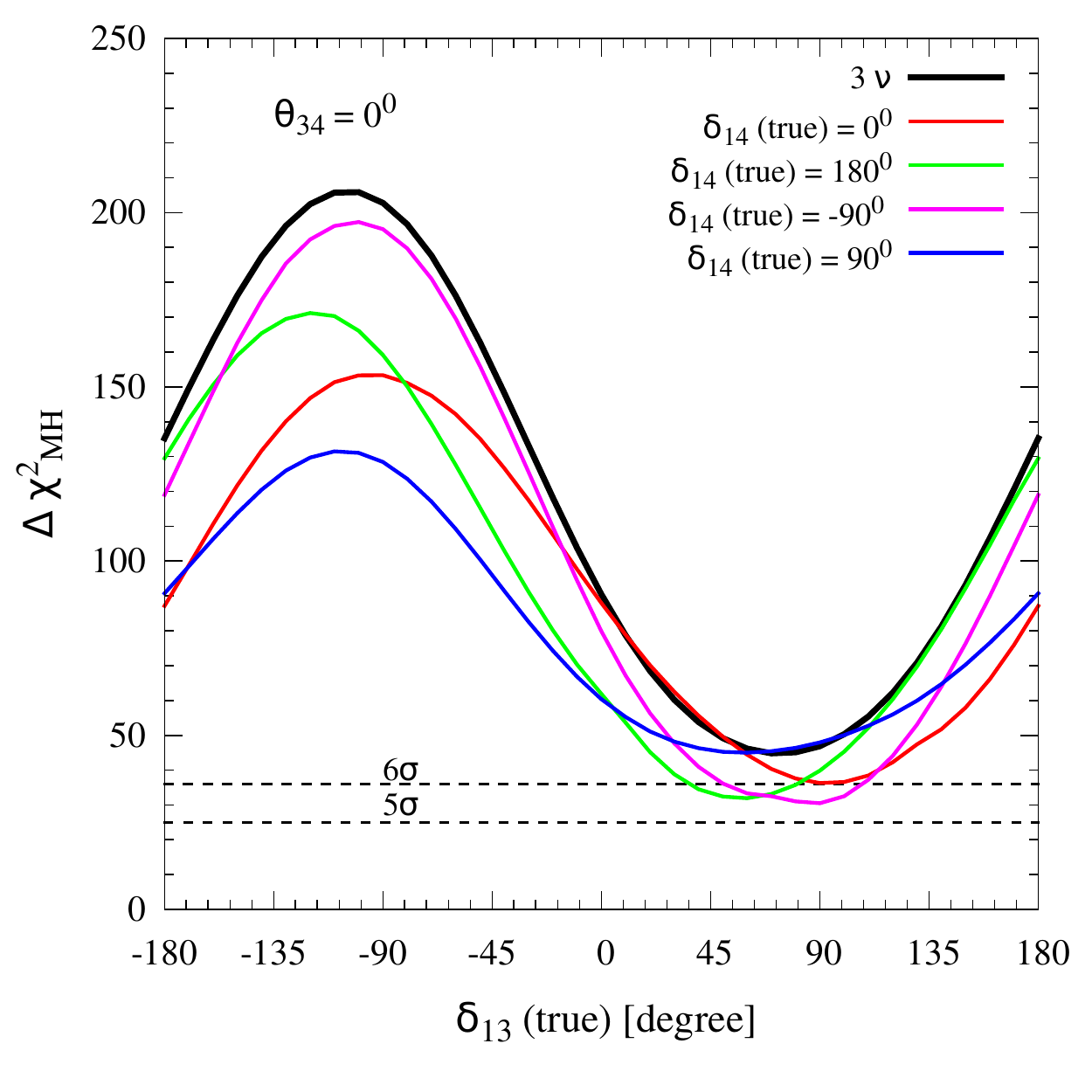}
\includegraphics[height=8. cm,width=8.0cm]{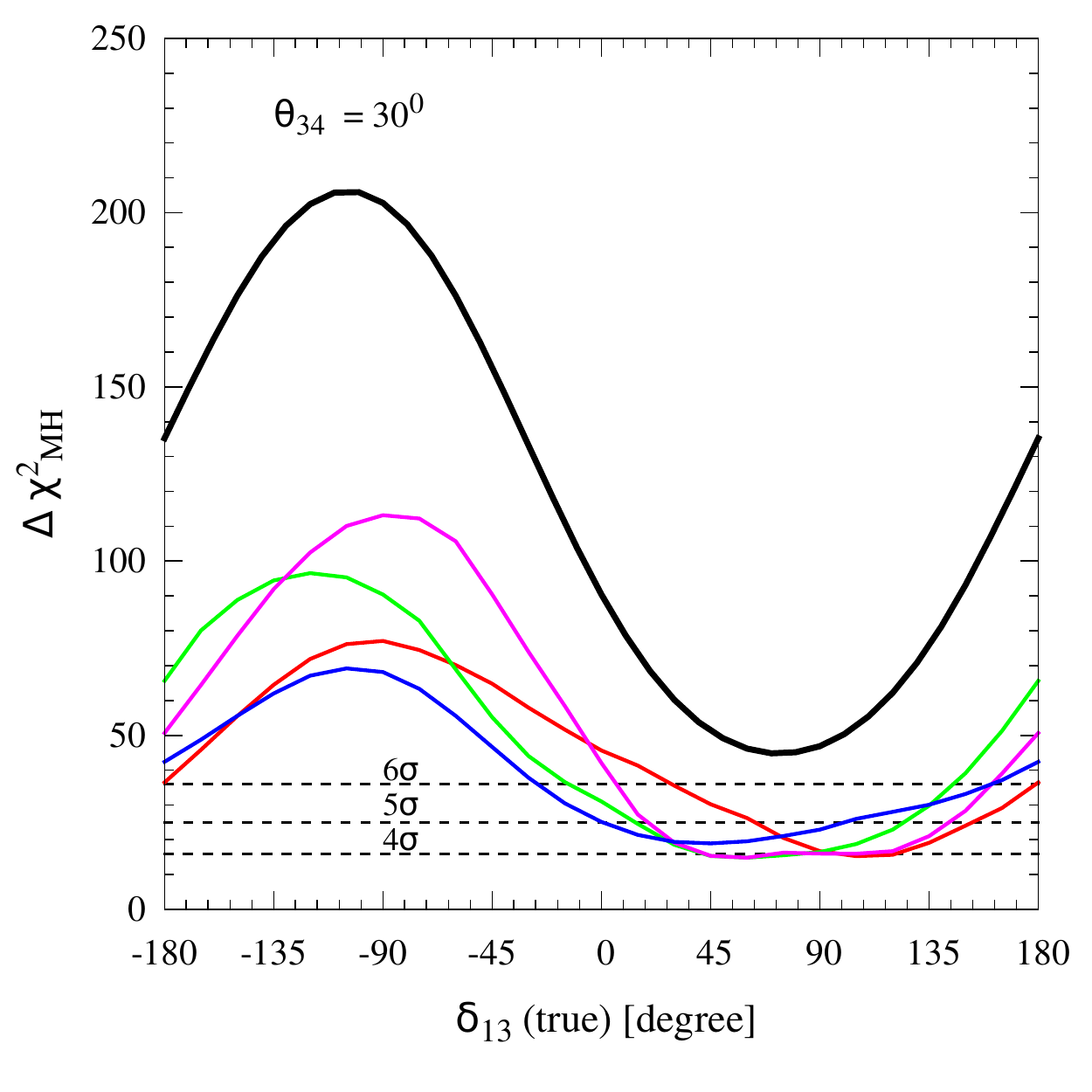}
}
\caption{Discovery potential for excluding the wrong hierarchy (IH) as a function 
of true $\delta_{13}$. In both panels, we have fixed true and test $\theta_{14} = \theta_{24} = 9^0$.
The left (right) panel refers to true and test $\theta_{34}$ = 0 ($30^0$). In each panel, 
we give the results for the 3-flavor case (black line), and for the 3+1 scheme for four different 
values of true $\delta_{14}$. In both the panels, we marginalize over test $\delta_{14}$ for the
3+1 scheme. On top of this, in the right panel, we also marginalize over true and test choices 
of the CP-phase $\delta_{34}$ over its full allowed range.}
\label{fig:mh}
\end{figure}

In our numerical analysis we calculate the DUNE discovery potential of the neutrino mass hierarchy. 
This is defined as the confidence level at which one can exclude the false (or wrong)
test hierarchy given a data set generated with the true hierarchy. We have taken the best
fit values of all the parameters as given in Table~\ref{tab:benchmark-parameters}.
In the 3-flavor scheme, we have marginalized over $\delta_{13}$ and $\theta_{23}$
within their full $3\sigma$ range. In the 3+1 scheme, we have also marginalized over 
the CP-phase $\delta_{14}$ (and  $\delta_{34}$ if $\theta_{34} \ne 0$). 
In Fig.~\ref {fig:mh}, we show the discovery potential of excluding the wrong hierarchy as a function
of the true value of $\delta_{13}$ considering NH as the true hierarchy choice. In each panel, we give 
the results for the 3-flavor case with the help of thick black curve. In both the panels, 
for the 3+1 scheme (colored curves), we have taken $\theta_{14} = \theta_{24} = 9^0$ while generating 
the data and also in the fit. In the left panel, we consider $\theta_{34} =0$ both in data and also in fit,
and give the results for four different values of the true $\delta_{14}$ ($-90^0$, $90^0$, $0^0$, and $180^0$) 
while marginalizing over test $\delta_{14}$ in its entire range of $[-\pi,\pi]$ in the fit. 
In the right panel, we show the same considering the true and test $\theta_{34} = 30^0$, and
therefore, we marginalize over both test $\delta_{14}$ and test $\delta_{34}$ in their entire ranges
of $[-\pi,\pi]$ in the fit. In the right panel, we also marginalize over true choices of unknown 
$\delta_{34}$ in its full range of $[-\pi,\pi]$ in the data.
In the left panel (corresponding to $\theta_{34} = 0$), we observe that the qualitative
behavior of the 3+1 curves are similar to the 3-flavor case. In particular, there is a 
maximum around $\delta_{13} \sim - 90^0$, and a minimum around $\delta_{13} \sim 90^0$. 
The opposite is true for the IH (not shown). It is evident that in general, there is a deterioration 
of the discovery potential for all values of the new CP-phase $\delta_{14}$. However, even 
in the region around the minimum, the sensitivity never drops below 5$\sigma$ confidence 
level. In the right panel (corresponding to $\theta_{34} = 30^0$), the situation is qualitatively
similar, but the sensitivity deteriorates more\footnote{We have checked that for the case 
$\theta_{34} = 9^0$, the results are intermediate between those found in the two cases 
$\theta_{34} = 0$ and $\theta_{34} = 30^0$ shown in Fig.~\ref {fig:mh}. In particular, 
the minimal sensitivity is approximately $5\sigma$.} compared to the left panel.
In particular, in the range $\delta_{13} \in [45^0, 135^0]$, the sensitivity can drop down to 4$\sigma$ 
confidence level. This range corresponds to the region of the space spanned by the thee CP-phases,
where there is basically a complete degeneracy at the level of the total number of events 
(in both neutrino and antineutrino channels), and the distinction between NH and IH is totally 
entrusted to the energy spectrum. A concrete example of this kind has been provided in the previous
subsection. To this regard, it is important to underline the fundamental difference between the 
experiments (like DUNE) that make use of an on-axis broad-band neutrino beam, and those 
using an off-axis configuration (T2K and NO$\nu$A). In this last case, there is basically no spectral 
information and, as a consequence, there are regions of the parameters space where the MH 
discovery potential is almost zero (see for example Fig.~14 in~\cite{Agarwalla:2016mrc}). 
In a nutshell, in the off-axis configuration, one has basically only the events counting at disposal,
while in the on-axis case, there is the extra information coming from the spectral shape.
Needless to say, the precondition to take advantage of this additional information is a good 
understanding of all the ingredients that enter the calculation of the event spectrum, and 
a refined treatment of the related systematic uncertainties. 
 
\section{CP-violation searches in the 3+1 scheme}
\label{sec:CPV}
   
In this section, we explore the impact of sterile neutrinos in the CPV searches of DUNE.
In the 3+1 scheme, there are three CP-phases, all of which can contribute to CPV. 
We first discuss the discovery potential of the CPV induced by the standard 3-flavor
CP-phase $\delta_{13} = \delta$. We will show that, in general, it tends to deteriorate 
in the 3+1 scheme with respect to the 3-flavor case. Then, we treat the sensitivity 
to the CPV induced by the other two CP-phases $\delta_{14}$ and $\delta_{34}$. 
Finally, we assess the capability of reconstructing the two phases $\delta_{13}$ 
and $\delta_{14}$.
 
\subsection{CP-violation discovery potential}

\begin{figure}[t!]
\centerline{
\includegraphics[height=7.6 cm,width=7.6cm]{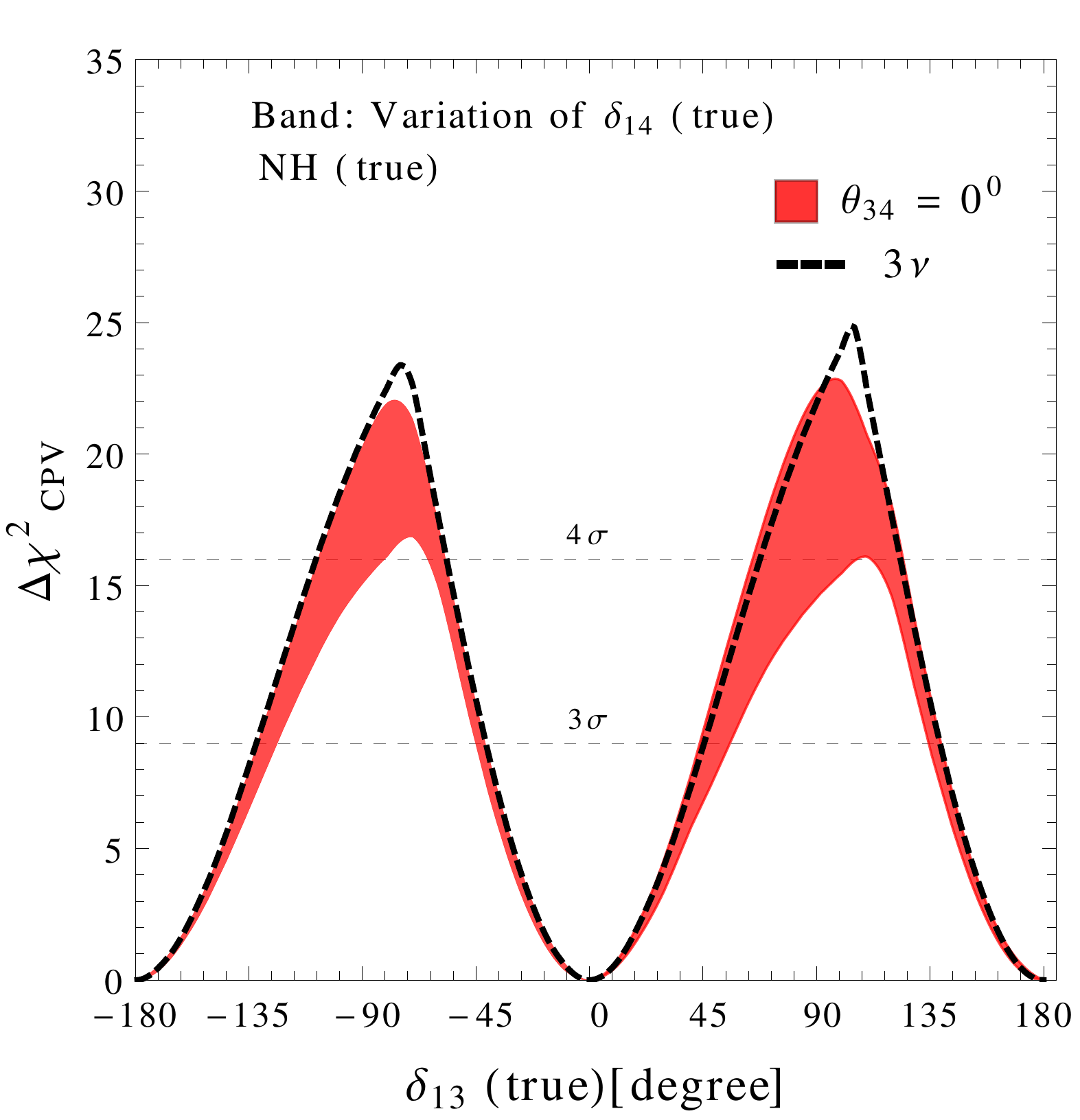}
\includegraphics[height=7.6 cm,width=7.6cm]{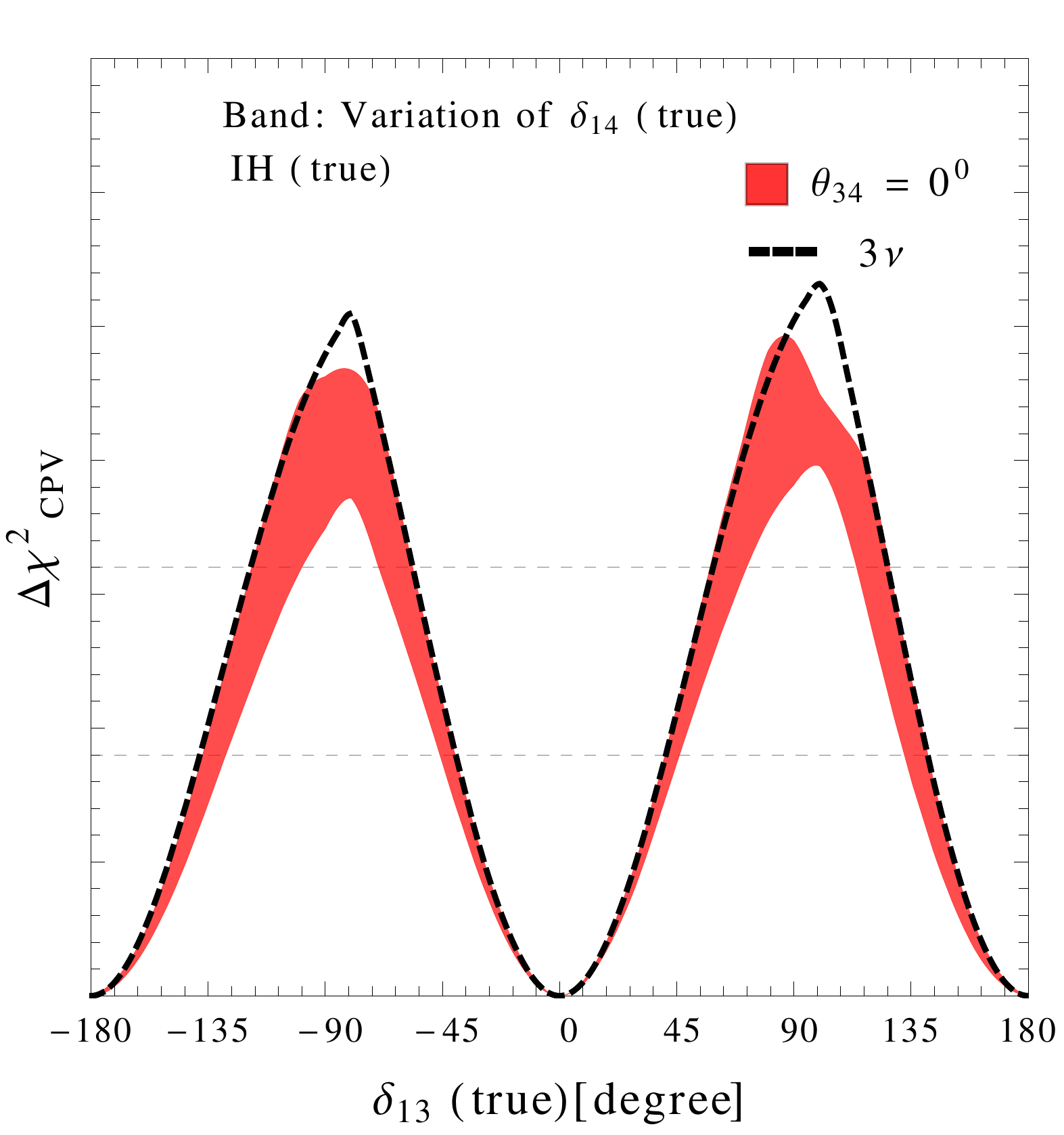}
} 
\caption{DUNE discovery potential of the CPV induced by $\delta_{13}$. The left (right) panel 
refers to true NH (IH). In both panels, the black dashed curve corresponds to the 3-flavor case. 
In 3+1 scenario, we fix the true and test values of $\theta_{14}$ = $\theta_{24}$ = $9^0$, 
and we take the true and test $\theta_{34}$ to be $0^0$ in both the panels. In both the panels, 
the red bands are obtained in the 3+1 scheme by varying the unknown true $\delta_{14}$ in 
its entire range of $[-\pi,\pi]$ in the data while marginalizing over test $\delta_{14}$ in the 
same range. Both in 3$\nu$ and 3+1 schemes, we marginalize over test $\theta_{23}$ 
and over both the choices of test hierarchy.
}
 \label{fig:CPV_3nu}
 \end{figure}
  
The discovery potential of CPV induced by a given true $\delta_{13}$ is defined 
as the confidence level at which one can reject the test hypothesis of no CPV i.e. 
the cases test $\delta_{13}=0$ and test $\delta_{13}=\pi$ in the fit from that given
true $\delta_{13}$. In Fig.~\ref{fig:CPV_3nu}, we report the discovery potential 
of CPV induced by $\delta_{13}$. We show it as a function of the true value of $\delta_{13}$. 
The left (right) panel refers to true NH (IH). As far as the standard oscillation parameters 
are concerned, we marginalize over test $\sin^2\theta_{23}$ in the range 0.34 to 0.68, 
and over both the choices of hierarchy in the fit for both 3$\nu$ and 3+1 cases. 
In both the panels, the black dashed curves correspond to the 3-flavor case. 
In 3+1 scenario, we fix the true and test values of $\theta_{14}$ and $\theta_{24}$ to be
$9^0$, and we take the true and test $\theta_{34}$ to be $0^0$ in both the panels, and 
therefore, $\delta_{34}$ becomes irrelevant. In both the panels, the red bands are obtained 
in the 3+1 scheme by varying the unknown true $\delta_{14}$ in its entire range of $[-\pi,\pi]$ 
in the data while marginalizing over test $\delta_{14}$ in the same range. 
Figure~\ref{fig:CPV_3nu} shows that the discovery potential can be deteriorated 
in the 3+1 scheme as compared to the 3-flavor framework. In particular, around 
$\delta_{13}$ $\sim$ $\pm$ $90^0$, where the discovery potential attains the maximal value, 
the discovery potential can decrease from $\sim$ 5$\sigma$ C.L. (3-flavor case) 
to $\sim$ 4$\sigma$ C.L. (3+1 case).

\begin{figure}[t!]
\centerline{
\includegraphics[width=1.2\textwidth]{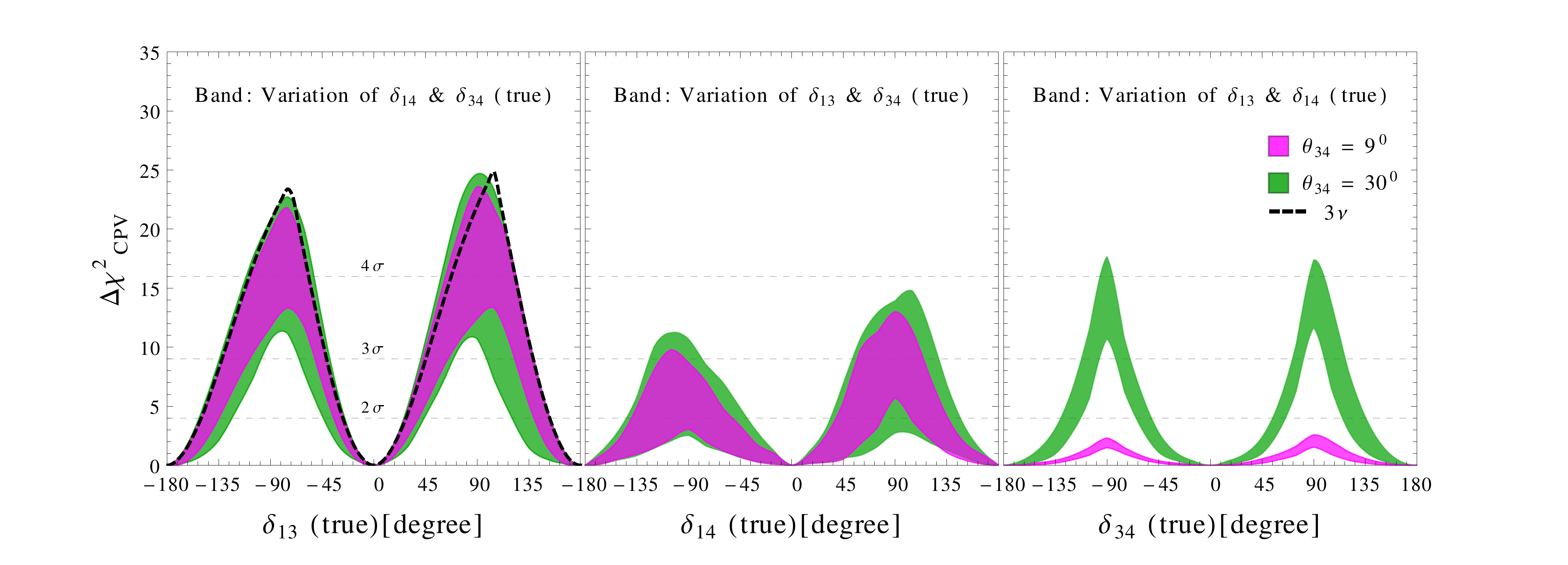}
}
\caption{
The bands displayed in the left, middle and right panels represent the discovery potential 
of the CPV induced, respectively, by $\delta_{13}$, $\delta_{14}$ and $\delta_{34}$ in the 
3+1 scheme. The thinner (magenta) bands correspond to the case in which all the three 
new mixing angles have the identical true and test value $\theta_{14} = \theta_{24} 
= \theta_{34} = 9^0$. The thicker (green) bands correspond to the case in which 
true and test $\theta_{14} = \theta_{24} = 9^0$, and true and test $\theta_{34} = 30^0$. 
In each panel, the bands have been obtained by varying the true values of
the two undisplayed CP-phases in their allowed ranges of $[-\pi,\pi]$ in the
data while marginalizing over their test values in the same range in the fit.
The left panel also reports the 3-flavor curve (black dashed line) for the sake 
of comparison. In all cases, we marginalize over hierarchy with NH as true choice.
}
\label{fig:CPV_4nu}
\end{figure}
  
\begin{figure}[t!]
\centerline{
\includegraphics[height=7.6 cm,width=7.6cm]{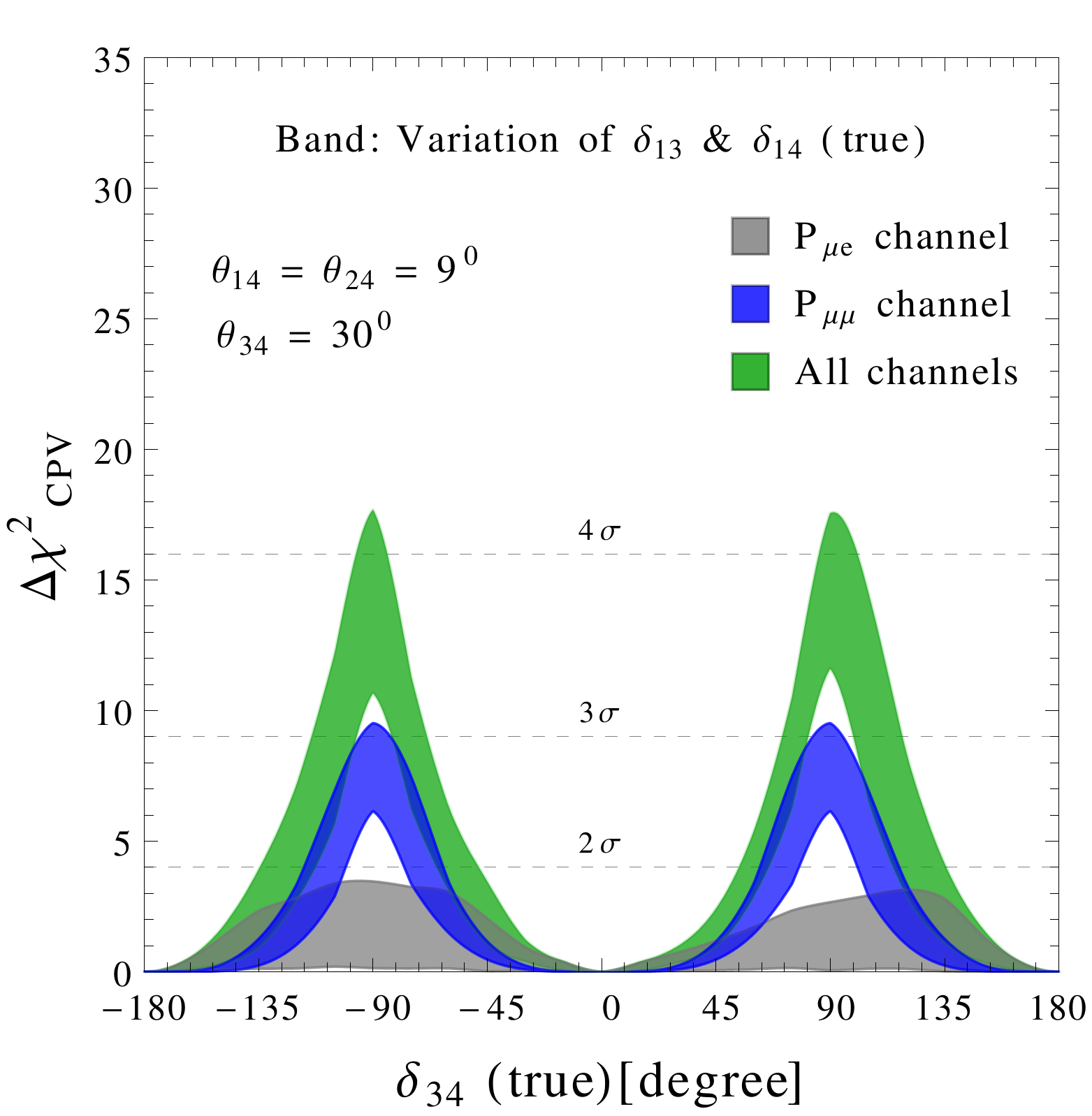}
} 
\caption{Partial contributions to the sensitivity to the CPV induced by the phase $\delta_{34}$ 
deriving from different channels and from their combination. The gray band is the contribution 
from $\nu_e$ and $\bar \nu_e $ appearance, the blue band refers to the $\nu_\mu$ and 
$\bar \nu_\mu$ disappearance, and the green band is the global sensitivity obtained 
including all channels. The three new mixing angles have been fixed to 
$\theta_{14} = \theta_{24} = 9^0$ and $\theta_{34} = 30^0$. The mixing angle $\theta_{23}$ 
has been fixed to be maximal (both true and test value). We marginalize over hierarchy with 
NH as true choice.}
\label{fig:CPV_split_34}
\end{figure}

In the 3+1 scheme, one expects CPV to come also from the two new phases 
$\delta_{14}$ and $\delta_{34}$. In the second and third panels of 
Fig.~\ref{fig:CPV_4nu}, we display the discovery potential of the CPV induced 
by such two phases. In the first panel, for the sake of comparison, we report 
the CPV discovery potential induced by the standard CP-phase $\delta_{13}$ 
so that one can have a global view of the sensitivities. The thinner (magenta) 
bands correspond to the case in which all the three new mixing angles have the 
identical true and test values $\theta_{14} = \theta_{24} = \theta_{34} = 9^0$. 
The thicker (green) bands correspond to the case in which true and test 
$\theta_{14} = \theta_{24} = 9^0$, and true and test $\theta_{34} = 30^0$.
In each panel, the bands have been obtained by varying the true values of
the two undisplayed CP-phases in their allowed ranges of $[-\pi,\pi]$ in the
data while marginalizing over their test values in the same range in the fit.
From the comparison of the three panels, we can see that if all the three mixing 
angles have the same value $\theta_{14} = \theta_{24} = \theta_{34} = 9^0$ 
(see the magenta bands), there is a clear hierarchy in the sensitivity to the 
three CP-phases. The standard phase $\delta_{13}$ comes first, $\delta_{14}$ 
comes next, and $\delta_{34}$ is the last one, inducing a negligible amount of CPV. 
In particular, we see that, in the less optimistic cases, corresponding to the 
lower border of the bands, only the standard CP-phase $\delta_{13}$ can give 
rise to a signal stronger than $3\sigma$ for an appreciable fraction of the true 
values of the phase. This fraction decreases if $\theta_{34}$ increases 
(compare the red band in the left panel of Fig.~\ref{fig:CPV_3nu} with the two bands 
in the left panel of Fig.~\ref{fig:CPV_4nu}). In Table~\ref{table:coverage}, for completeness, 
we report such a fraction for the three values of $\theta_{34} = 0^0, 9^0$, and $30^0$ 
as well as for the 3-flavor case. In the same table, we also report as a benchmark 
the ``guaranteed'' discovery potential for the particular value $\delta_{13} = -90^0$. 
Concerning the second CP-phase $\delta_{14}$ (see the middle panel), the discovery 
potential spans over a wide band. In favorable cases, one may observe a signal above 
the 3$\sigma$ level. However, the lower border of the band is always below the $\sim$ 
2$\sigma$ confidence level. This implies that there is not a guaranteed discovery 
potential at the 3$\sigma$ confidence level for such a phase. Finally, let us come to the 
third CP-phase $\delta_{34}$, shown in the right panel. We see that if the mixing angle 
$\theta_{34}$ is very big, the sensitivity can be appreciable. Also, we note that the shape 
of the band is rather different from that obtained for the other two CP-phases. 
This different behavior can be understood by observing that in this case, the 
$\nu_\mu \to \nu_\mu$ disappearance channel also contributes to the sensitivity. 
In fact, it is well known that the $\nu_\mu$ survival probability has a pronounced sensitivity 
to the NSI-like coupling $\varepsilon_{\mu \tau}$. From the expression of the Hamiltonian 
in Eq.~(\ref{eq:Hdyn_2}), one can see that $\varepsilon_{\mu \tau} = r  s_{24}\tilde s_{34}^*$, 
hence a sensitivity to the $\nu_\mu$ survival probability to the CP-phase $\delta_{34}$ is expected%
\footnote{It should be noted that the $\nu_\mu$ survival probability depends only on the 
real part of the new dynamical NSI-like coupling 
${\mathrm  {Re}}(\varepsilon_{\mu \tau}) = {\mathrm  {Re}} (r  s_{24}\tilde s_{34}^*) = r s_{24} s_{34} \cos \delta_{34}$. 
This implies that the sensitivity to $\delta_{34}$ is attained only via terms proportional to $\cos \delta_{34}$. 
This feature is reflected in the particular shape of the blue band in Fig.~\ref{fig:CPV_split_34}, 
which is symmetrical under a reflection around $\delta _{34} = 0^0$, i.e., under the transformation 
$\delta_{34} \to - \delta_{34}$. It is worth noticing that while the $\cos \delta_{34}$ term does not give 
rise to manifest CPV, its measurement allows one to establish the existence of CPV indirectly.}.
In order to make this point more clear, in Fig.~\ref{fig:CPV_split_34} we display the partial contributions 
to the CPV induced by $\delta_{34}$ deriving from the $\nu_\mu \to \nu_e$ channel, 
the $\nu_\mu \to \nu_\mu$ channel and from their combination. In this plot we have fixed the 
test value of $\theta_{23}$ at its true (maximal) value because the $\nu_\mu \to \nu_e$ channel, 
when taken alone, has a limited sensitivity to this parameter. Finally, we remark that the 
$\nu_\mu \to \nu_\mu$ channel has almost no role in the sensitivity to the other two CP-phases 
$\delta_{13}$ and $\delta_{14}$, as we have explicitly verified numerically.

\begin{table}[t!]
{%
\newcommand{\mc}[3]{\multicolumn{#1}{#2}{#3}}
\newcommand{\mr}[3]{\multirow{#1}{#2}{#3}}
\begin{center}
\begin{tabular}{|c|c|c|c|}
\hline
& $\theta_{34}$ & ${\mathrm N} \sigma_{min}$ \,[$\delta_{13} (\rm true)$ = -$90^0$] & CPV coverage (3$\sigma$) \\
\hline
3$\nu$ &    & 4.5 &    50.0\%\\
\hline
\mr{3}{*}{3+1} & $0^0$ & 3.9  & 43.2\% \\
\cline{2-4}
&  $9^0$ & 3.4  &  32.0\%  \\
\cline{2-4}
&  $30^0$ & 3.3  & 16.0\%  \\
\hline
\end{tabular}
\end{center}
}%
\caption{Discovery potential and coverage for the CPV induced by $\delta_{13}$ 
for four benchmark models in the NH case. The first column reports the scheme 
under consideration. The second column reports the value of $\theta_{34}$ 
(not relevant for the 3$\nu$ scheme). The third column reports the discovery potential 
(in units of standard deviations) for the particular value $\delta_{13} = -90^0$.
The fourth column reports the coverage at the $3\sigma$ level. Note that 
in the 3+1 scheme, the figures reported in the third (fourth) columns represent 
the minimal ``guaranteed'' discovery potential (coverage) as derivable from the 
lower border of the corresponding band in the left panel of 
Figs.~\ref{fig:CPV_3nu} and~\ref{fig:CPV_4nu}.}
\label{table:coverage}
\end{table}

\subsection{Reconstruction of the CP-phases}

\begin{figure}[t!]
\centerline{
\includegraphics[height=8. cm,width=8.cm]{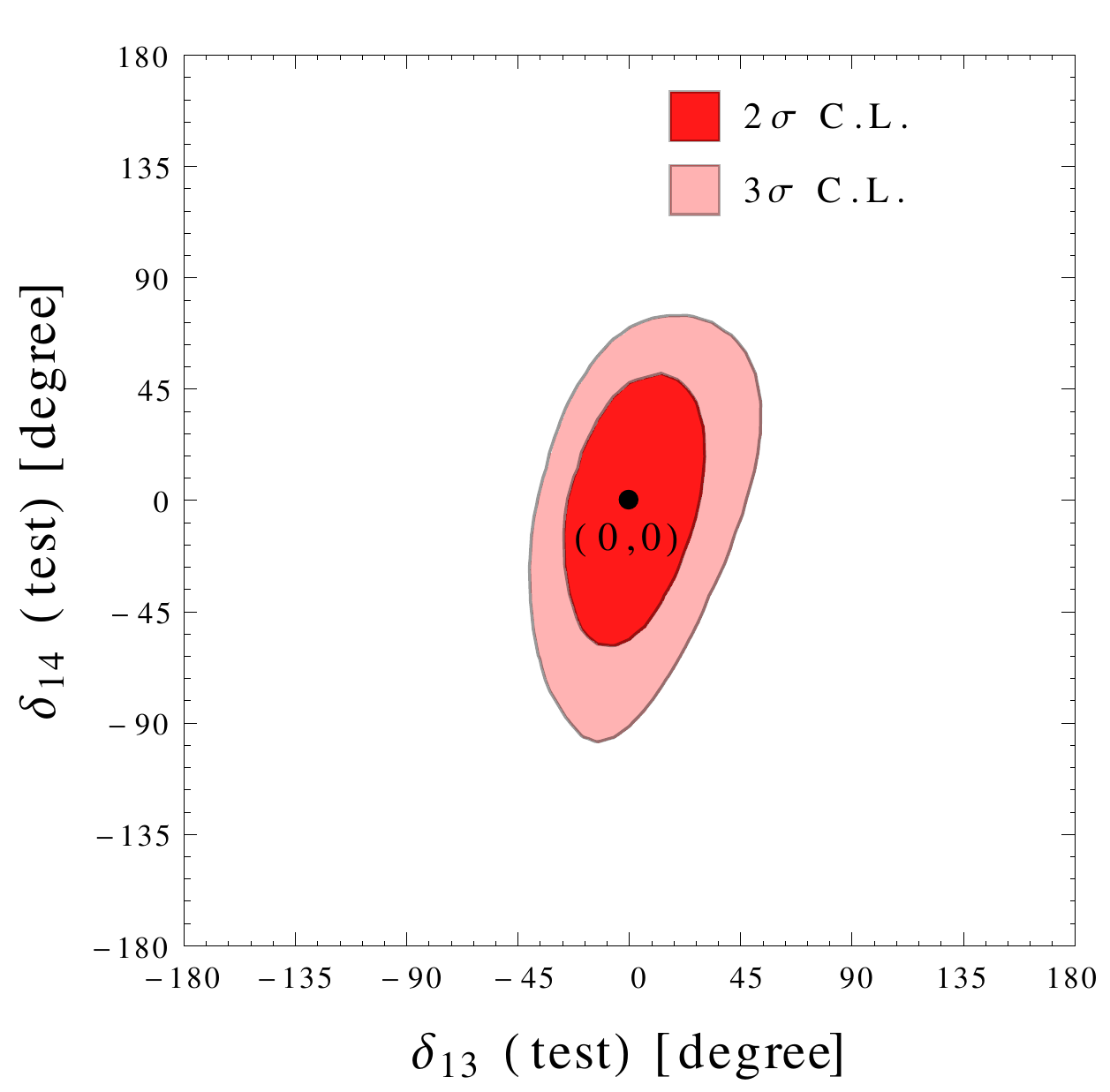}
\includegraphics[height=8. cm,width=8.cm]{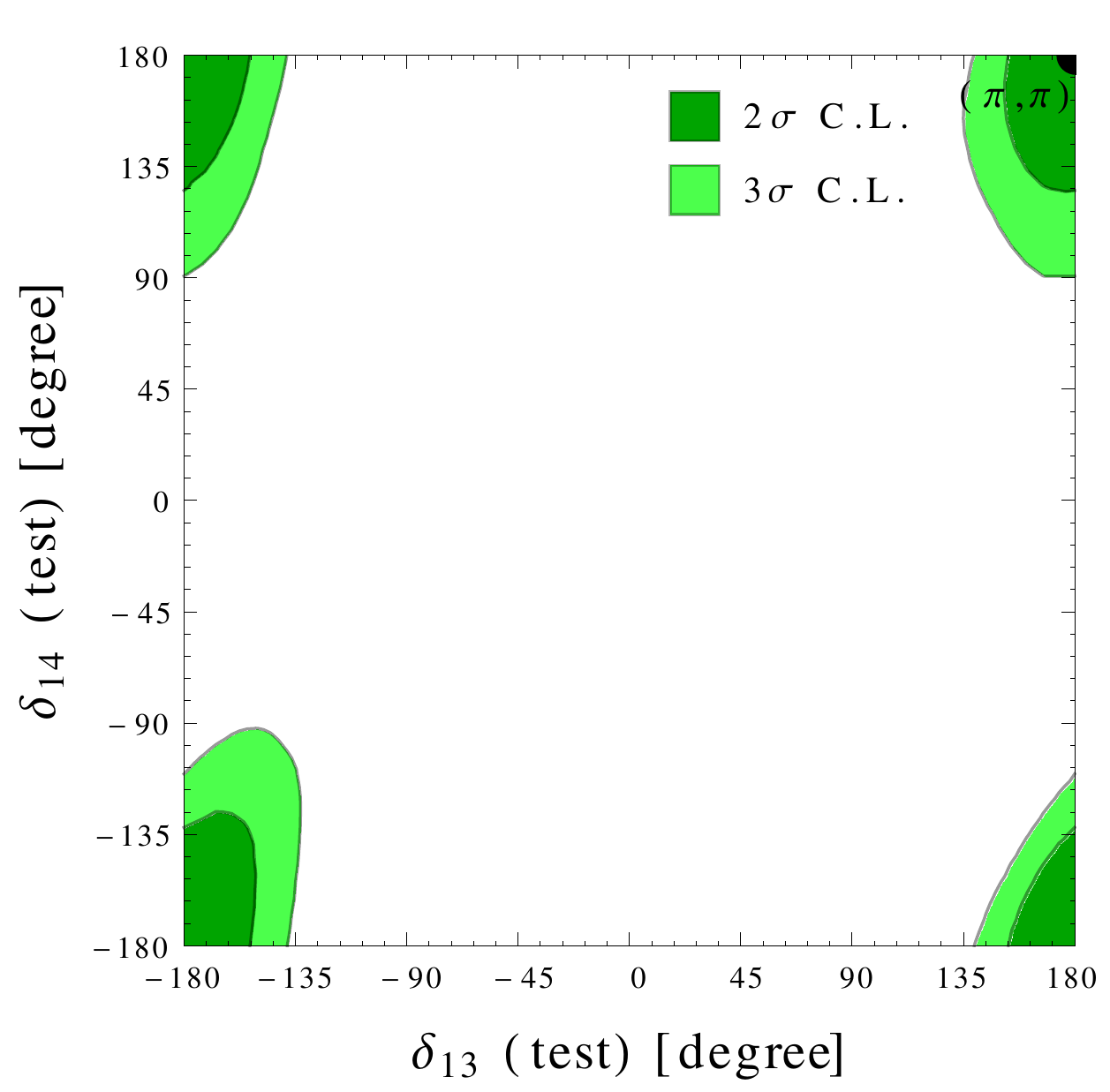}
}
\centerline{
\includegraphics[height=8. cm,width=8cm]{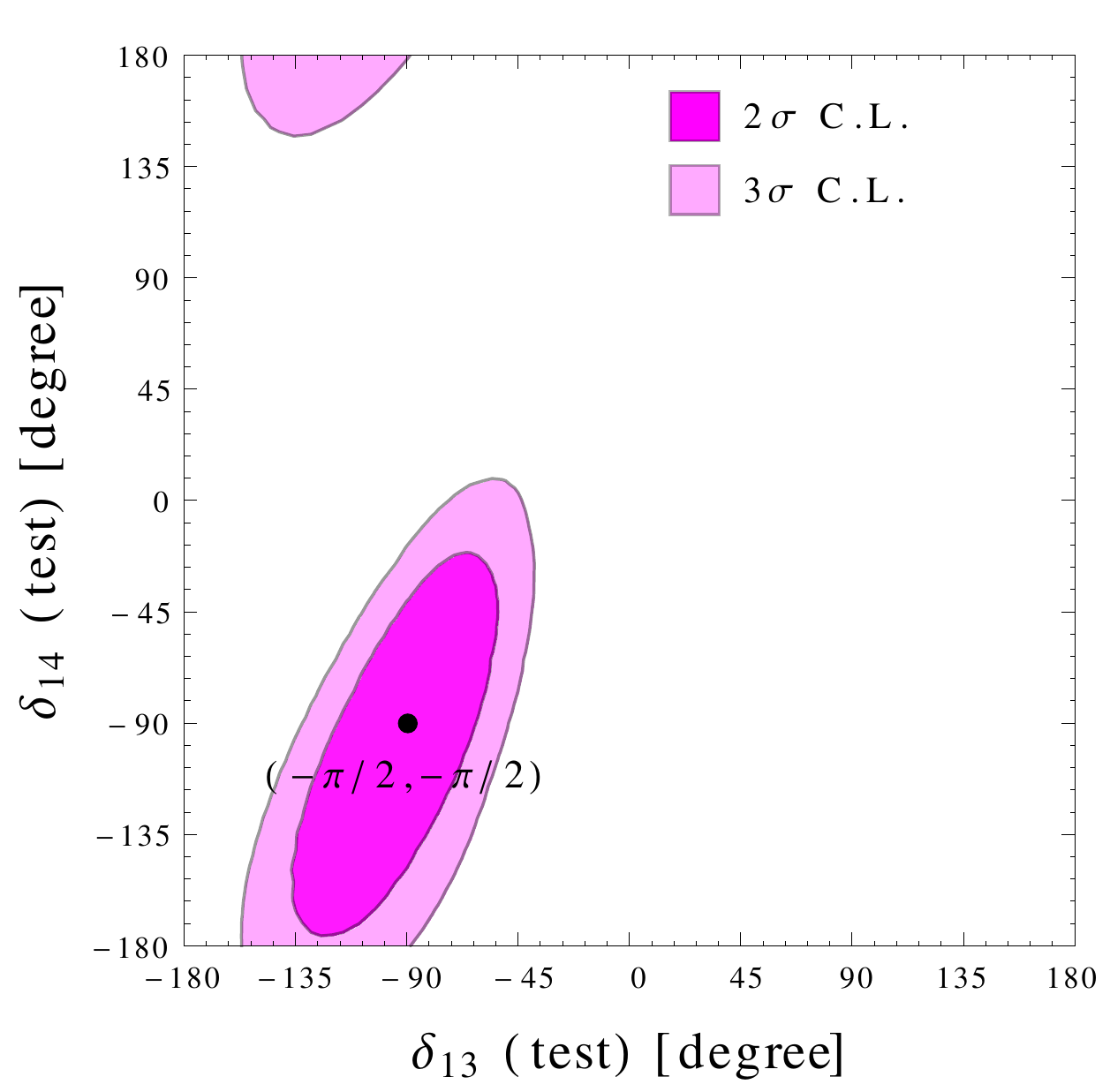}
\includegraphics[height=8. cm,width=8cm]{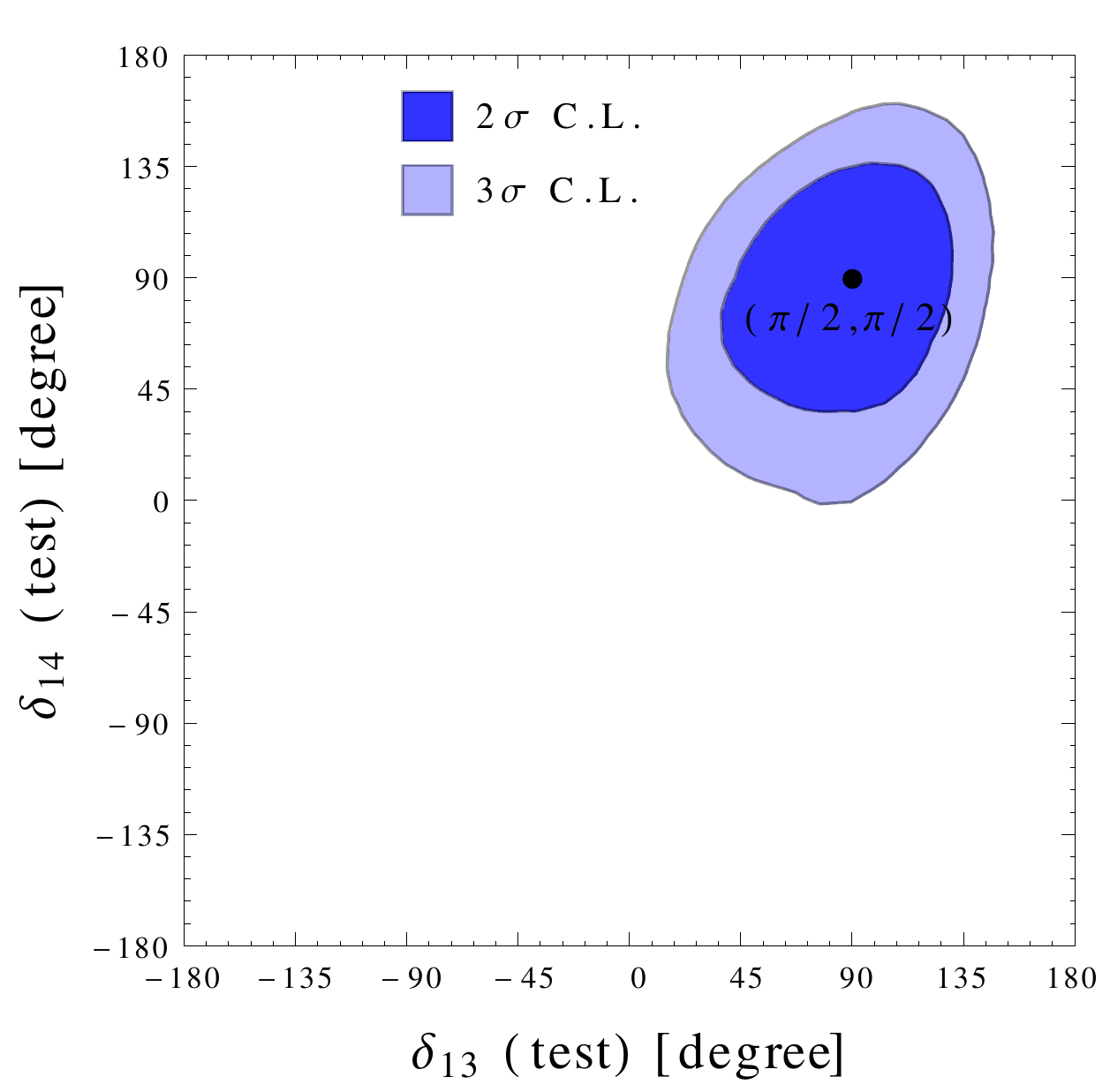}
}
\caption{Reconstructed regions for the two CP-phases $\delta_{13}$ and $\delta_{14}$ 
for the four choices of their true values indicated in each panel. The NH is taken as the 
true hierarchy, while we have marginalized over the two possible hierarchies in the test model. 
The contours refer to 2$\sigma$ and 3$\sigma$ levels. We have fixed the values 
$\theta_{34}$ (true) = $0^0$.}
\label{fig:CPV_rec_th34_0}
\end{figure}

\begin{figure}[t!]
\centerline{
\includegraphics[height=8. cm,width=8.cm]{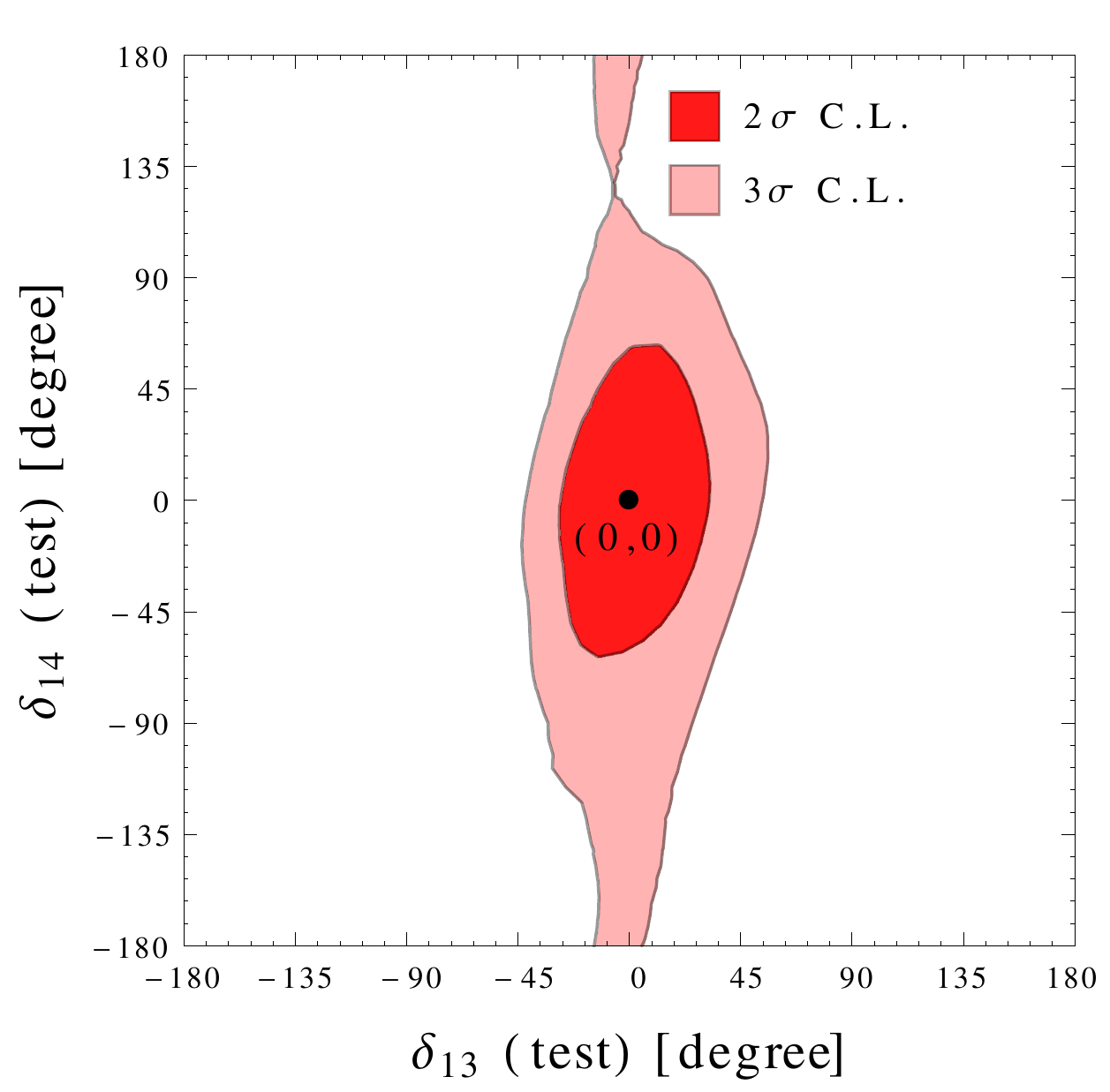}
\includegraphics[height=8. cm,width=8.cm]{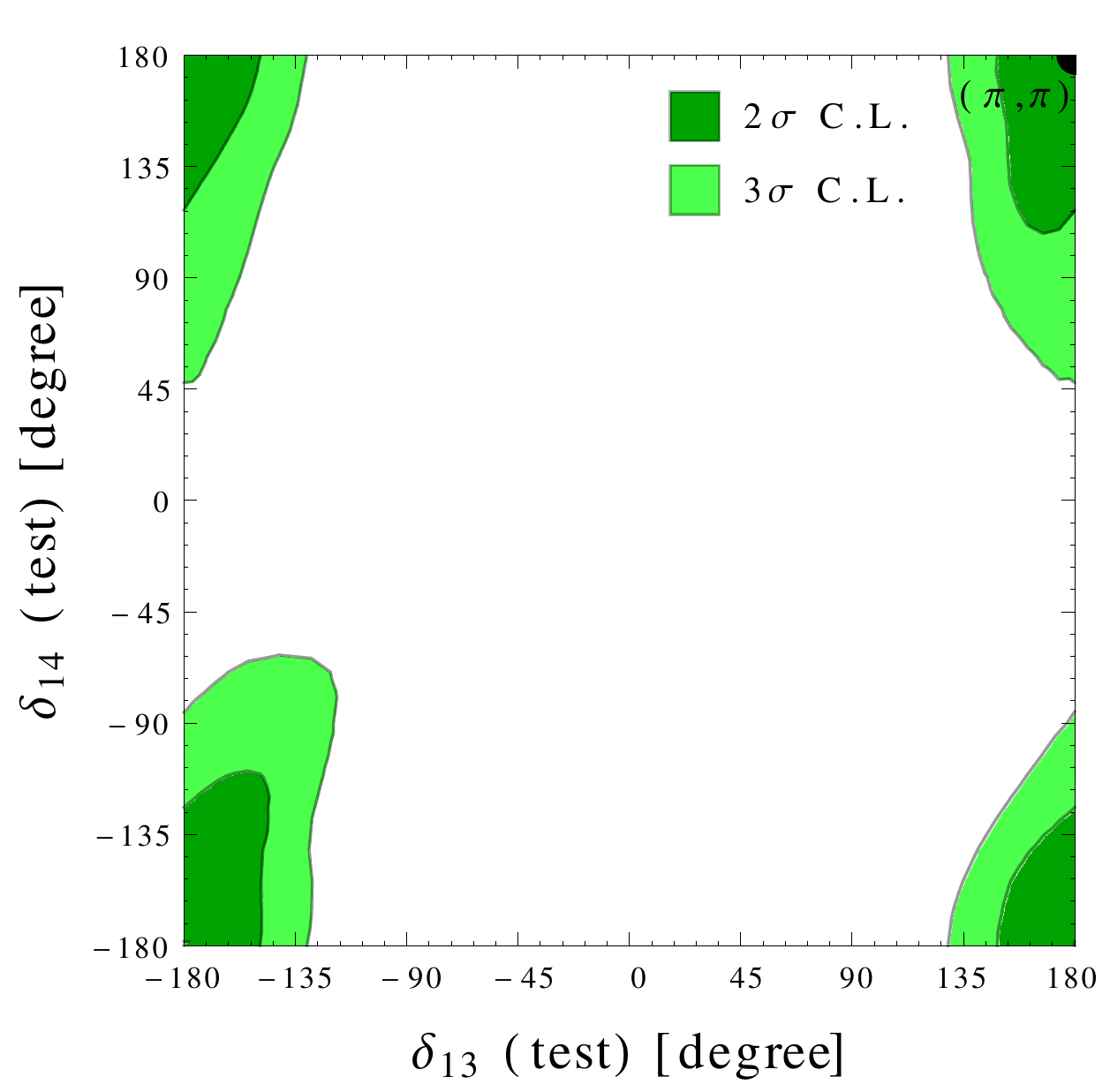}
}
\centerline{
\includegraphics[height=8. cm,width=8cm]{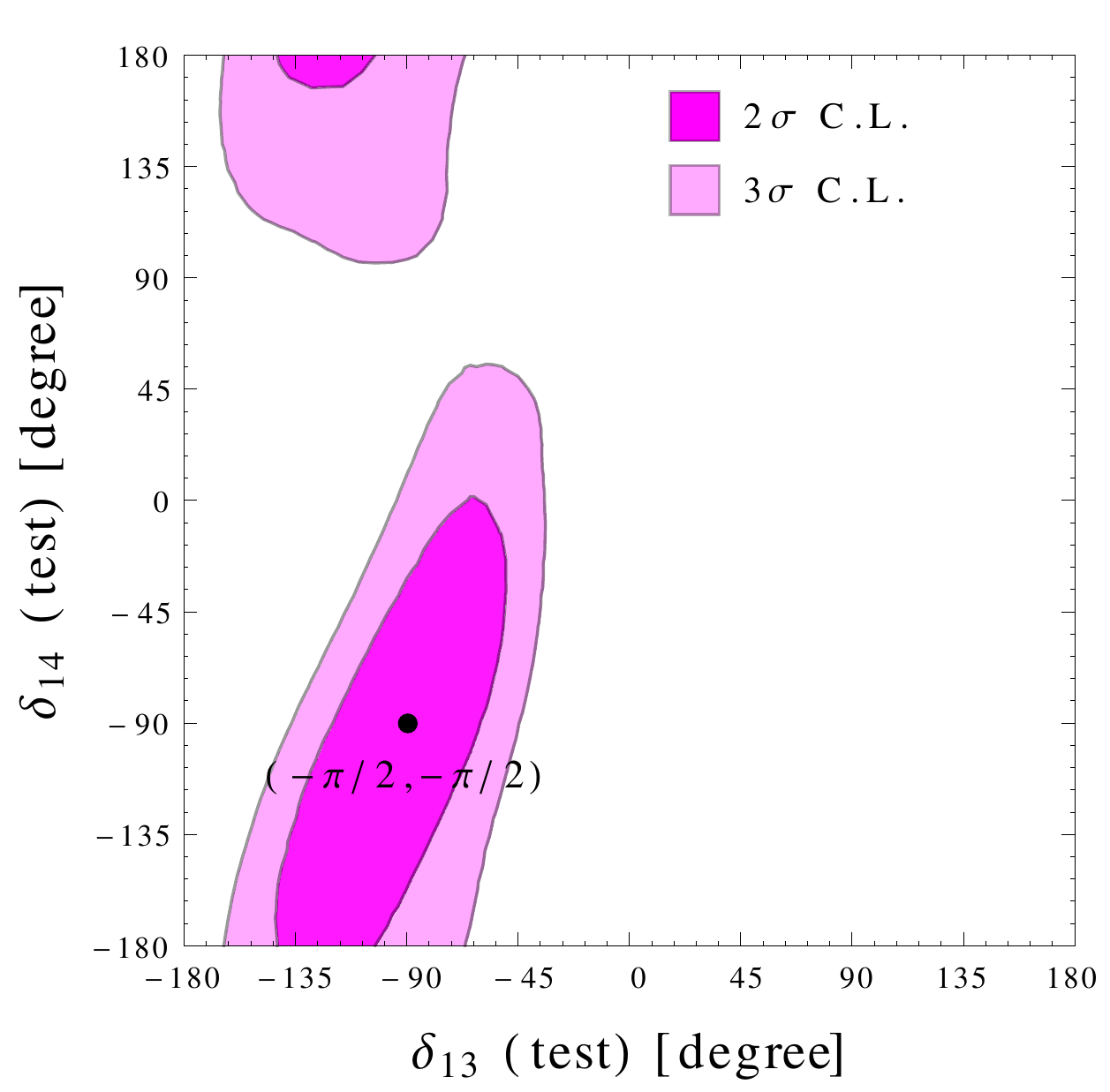}
\includegraphics[height=8. cm,width=8cm]{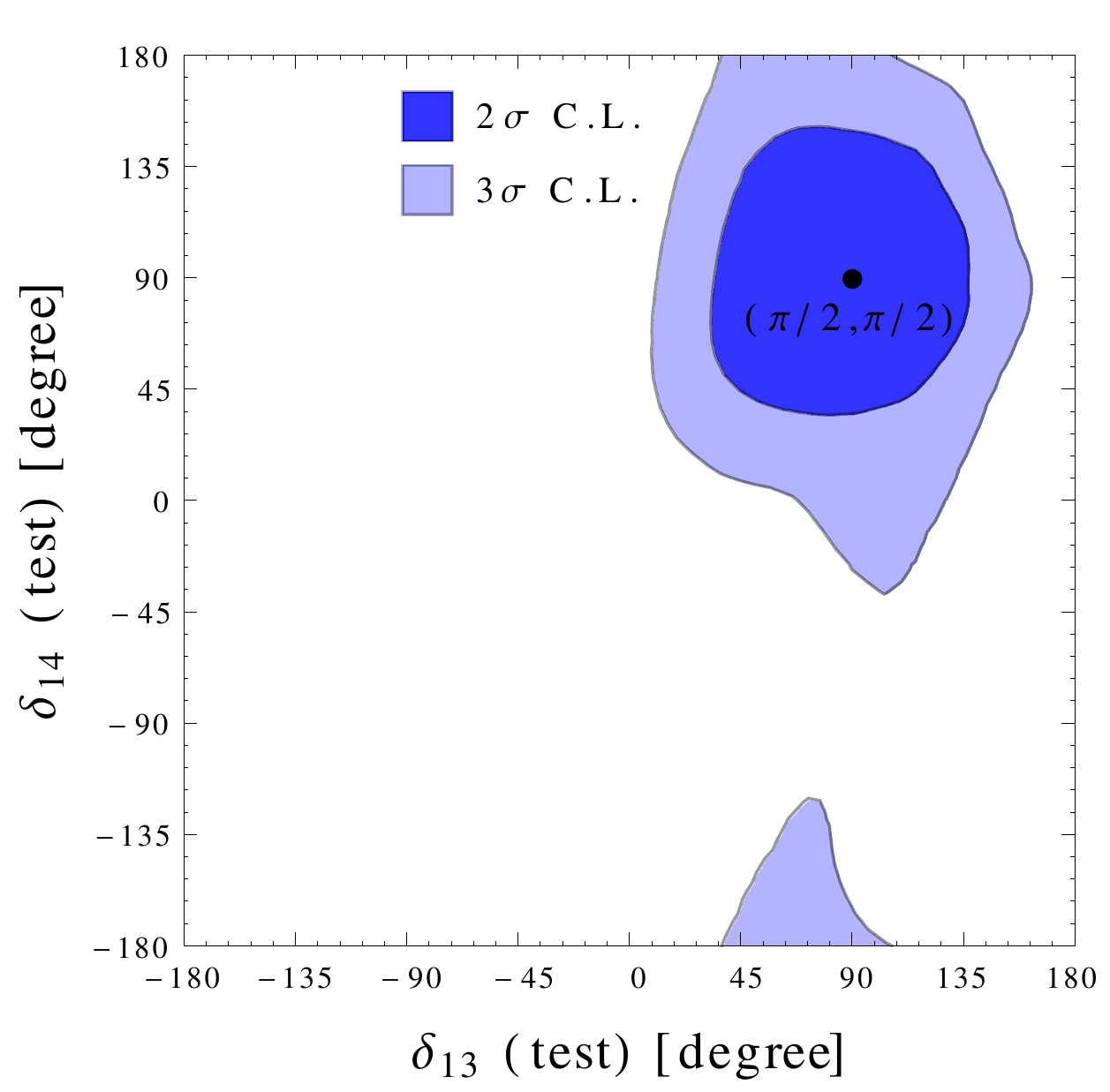} 
}
\caption{Reconstructed regions for the two CP-phases $\delta_{13}$ and $\delta_{14}$ 
for the four choices of their true values indicated in each panel. The NH is taken as the 
true hierarchy, while we have marginalized over the two possible hierarchies
in the test model. The contours refer to 2$\sigma$ and 3$\sigma$ confidence levels. 
We have fixed $\theta_{34}$ (true) = $30^0$, and have marginalized over 
$\delta_{34}$ (both true and test values) over its range of variability.} 
\label{fig:CPV_rec_th34_30}
\end{figure}

In the previous subsection, we have focused our attention on the discovery potential 
of the CPV arising from the three CP-phases involved in the 3+1 scheme.
However, we deem it important to pursue the exploration of the CP-phases 
independently of the amount of CPV (if any) that they may generate. In principle, 
any value of the CP-phases is plausible, including the CP-conserving cases.
According to this point of view, one should try to answer the following question:
what is the capability of reconstructing the true values of the CP-phases. 
In the following, we address this issue, confining our study to the two 
CP-phases $\delta_{13}$ and $\delta_{14}$, to which one expects to have 
a more pronounced sensitivity. 

In Fig.~\ref{fig:CPV_rec_th34_0}, we have performed this exercise under
the assumption that true and test values of $\theta_{34} =0$, and therefore,
the third CP-phase $\delta_{34}$ becomes irrelevant. In each panel, 
we show the regions reconstructed around the representative true values 
of $\delta_{13}$ and $\delta_{14}$. In all cases, we 
have taken the NH as the true hierarchy in the data, and then marginalized over 
both NH and IH in theory. However, we find that the marginalization procedure 
never selects the wrong hierarchy. Hence, there is no degeneracy between 
mass hierarchy and the CP-phases. Similar results (not shown) are obtained 
for the IH case. The two upper panels refer to two representative CP-conserving 
scenarios $[0,0]$ and $[\pi,\pi]$. The third and fourth panels refer to two representative 
CP-violating cases $[-\pi/2, -\pi/2]$ and $[\pi/2, \pi/2]$. The two confidence levels 
correspond to 2$\sigma$ and 3$\sigma$ (1 d.o.f.). We see that in all cases we obtain 
a unique reconstructed region at the 3$\sigma$ level%
\footnote{Note that this is true also in the second panel, because the 
four corners of the square form a connected region due to the cyclic 
nature of the two CP-phases.}.
The typical 1$\sigma$ level uncertainty on the reconstructed CP-phases 
is approximately $20^0$ ($30^0$) for $\delta_{13}$ ($\delta_{14}$). 
The regions in Fig.~\ref{fig:CPV_rec_th34_0} should be compared with 
the analogous ones obtained with the combination of the prospective 
data of T2K and NO$\nu$A (see Fig.~13 in~\cite{Agarwalla:2016mrc}).
From such a comparison, one can see that DUNE is much more effective 
in the reconstruction of the CP-phases. In fact, in the combination of 
T2K and NO$\nu$A, a good reconstruction is possible only at the 
1$\sigma$ confidence level, and in favorable cases, at the 2$\sigma$ level. 
At higher confidence levels, the reconstruction capability basically disappears. 
In DUNE, the situation is much improved. This is imputable both to the higher 
statistics and to the spectral shape information attainable from 
the wide-band neutrino flux.

We close our discussion on the CP-phases by showing the impact of $\theta_{34}$
and of the ignorance of the related CP-phase $\delta_{34}$ on the reconstruction of
the two CP-phases $\delta_{13}$ and $\delta_{14}$. Figure~\ref{fig:CPV_rec_th34_30}
is analogous to Fig.~\ref{fig:CPV_rec_th34_0}, but now, we assume $\theta_{34} = 30^0$
both in data and in theory. In this case, the transition probability depends also on the 
third CP-phase $\delta_{34}$. Since it is unknown, one has to marginalize over 
the true and test values of this CP-phase while reconstructing the other two phases. 
As evident from all panels of Fig.~\ref{fig:CPV_rec_th34_30}, the quality of the 
reconstruction deteriorates. In particular, the uncertainty on $\delta_{14}$
increases roughly by two times (from $\sim 30^0$ to $\sim 60^0$), while that 
on $\delta_{13}$ remains basically unchanged (still around $\sim 25^0$). 
Hence, the reconstruction of the standard CP-phase $\delta_{13}$ is more robust 
than that of new CP-phase $\delta_{14}$ with respect to the perturbations induced 
by a large value of the mixing angle $\theta_{34}$. This behavior can be traced 
to the fact that, differently from $\delta_{13}$, the two new CP-phases $\delta_{14}$ 
and $\delta_{34}$ both enter at the dynamical level 
[see Eqs.~(\ref{eq:Hdyn_2}-\ref{eq:Hdyn_3})], and in particular, in the NSI-like 
matrix element $\varepsilon_{e\tau} \propto \tilde s_{14} \tilde s_{34}^*$. 
Therefore, one may expect a certain degree of degeneracy among the 
two new CP-phases, which in the marginalization process over the 
undisplayed CP-phase $\delta_{34}$, translates into a deterioration 
in the reconstruction of $\delta_{14}$.

\section{Conclusions and Outlook}
\label{Conclusions}

We have investigated the impact of one light eV scale sterile neutrino 
on the prospective data expected to be collected at the planned 
long-baseline experiment DUNE. We have found that the discovery potential 
of the neutrino mass hierarchy (MH), remains above 5$\sigma$ confidence level 
if all the three new mixing angles are relatively small 
($\theta_{14} = \theta_{24} = \theta_{34} = 9^0$). In contrast, if the third 
mixing angle $\theta_{34}$ is taken at its upper limit ($\theta_{34} = 30^0$), 
the MH sensitivity can drop to 4$\sigma$ confidence level. Our analysis 
has clearly shown that the spectral information attainable from the 
wide-band spectrum employed by DUNE is crucial to preserve a 
good sensitivity to the MH also in the 3+1 scheme. We have also assessed 
the sensitivity to the CPV induced both by the standard CP-phase 
$\delta_{13} \equiv \delta$, and by the new CP-phases 
($\delta_{14}$ and $\delta_{34}$ in our parametrization). 
We have found that the performance of DUNE in claiming the discovery
of CPV induced by $\delta_{13}$ gets deteriorated as compared to the
3-flavor case. In particular, the maximal sensitivity 
(reached around $\delta_{13}$ $\sim$ $\pm$ $90^0$) decreases from
$5\sigma$ to $4\sigma$ confidence level if all the three new mixing angles 
are small ($\theta_{14} = \theta_{24} = \theta_{34} = 9^0$), and can drop 
almost to $3\sigma$ confidence level if $\theta_{34} = 30^0$. 
The sensitivity to the CPV induced by the new CP-phase $\delta_{14}$ 
can reach 3$\sigma$ C.L. for an appreciable fraction of its true values,
but never reaches 4$\sigma$ confidence level. The sensitivity to the 
third CP-phase $\delta_{34}$, which arises exclusively through matter effects, 
is appreciable only if $\theta_{34}$ is large. Interestingly, we have found that
the sensitivity to $\delta_{34}$ stems from both the $\nu_e$ appearance 
and $\nu_\mu$ disappearance channels. We have finally investigated 
the capability of DUNE in reconstructing the true values of the two 
CP-phases $\delta_{13}$ and $\delta_{14}$. The typical 1$\sigma$ level 
uncertainty on the reconstructed CP-phase $\delta_{13}$ ($\delta_{14}$)
is approximately $20^0$ ($30^0$) provided $\theta_{34} =0$.
But, in case of large $\theta_{34}$, the reconstruction of $\delta_{14}$ 
(but not that of $\delta_{13}$) becomes poor. So, finally, we can 
conclude that the results presented in this paper clearly demonstrates that 
in the presence of a light eV-scale sterile neutrino, the proposed 
LBL experiments such as DUNE would be quite sensitive to the 
new CP-phases associated to the sterile state, and therefore, 
would play a complementary role to the SBL experiments.

\subsubsection*{Acknowledgments}

S.K.A. is supported by the DST/INSPIRE Research Grant [IFA-PH-12],
Department of Science \& Technology, India. A.P. is supported by the 
Grant ``Future In Research'' {\it Beyond three neutrino families},
contract no. YVI3ST4, of Regione Puglia, Italy. 

\bibliographystyle{JHEP}
\bibliography{Sterile-References}

\end{document}